\RequirePackage{fix-cm}
\documentclass{svjour31}                     
\RequirePackage{hyperref}
\smartqed  
\usepackage{graphicx}

\usepackage{amsthm,amsmath}
\usepackage{booktabs}
\usepackage{multirow}
\usepackage{natbib}
\usepackage{algorithm}
\usepackage{rotating}
\usepackage{algpseudocode}
\usepackage[dvipsnames]{xcolor}
\usepackage{caption}
\usepackage{subcaption}
\usepackage{lineno}

\usepackage{geometry}
\journalname{Journal}
\begin{document}

\title{Convolutional Neural Networks based Focal Loss for Class Imbalance Problem: A Case Study of Canine Red Blood Cells Morphology Classification
}

\titlerunning{CNNs based Focal Loss for Class Imbalance Problem: Canine Red Blood Cells Morphology Classification}        

\author{Kitsuchart Pasupa \and
        Supawit Vatathanavaro \and
        Suchat Tungjitnob
}


\institute{K. Pasupa (Corresponding Author) \at
              Faculty of Information Technology, King Mongkut's Institute of Technology Ladkrabang 10520, Bangkok, Thailand \\
              \email{kitsuchart@it.kmitl.ac.th}
           \and
           S. Vatathanavaro \at
              Faculty of Information Technology, King Mongkut's Institute of Technology Ladkrabang 10520, Bangkok, Thailand \\
              \email{62606061@kmitl.ac.th}           
           \and
           S. Tungjitnob \at
              Faculty of Information Technology, King Mongkut's Institute of Technology Ladkrabang 10520, Bangkok, Thailand \\
              \email{62606062@kmitl.ac.th}
}


\maketitle
\begin{abstract}
Morphologies of red blood cells are normally interpreted by a pathologist. It is time-consuming and laborious. Furthermore, a misclassified red blood cell morphology will lead to false disease diagnosis and improper treatment. Thus, a decent pathologist must truly be an expert in classifying red blood cell morphology. In the past decade, many approaches have been proposed for classifying human red blood cell morphology. However, those approaches have not addressed the class imbalance problem in classification. A class imbalance problem---a problem where the numbers of samples in classes are very different---is one of the problems that can lead to a biased model towards the majority class. Due to the rarity of every type of abnormal blood cell morphology, the data from the collection process are usually imbalanced. In this study, we aimed to solve this problem specifically for classification of dog red blood cell morphology by using a Convolutional Neural Network (CNN)---a well-known deep learning technique---in conjunction with a focal loss function, adept at handling class imbalance problem. The proposed technique was conducted on a well-designed framework: two different CNNs were used to verify the effectiveness of the focal loss function and the optimal hyper-parameters were determined by 5-fold cross-validation. The experimental results show that both CNNs models augmented with the focal loss function achieved higher $F_{1}$-scores, compared to the models augmented with a conventional cross-entropy loss function that does not address class imbalance problem. In other words, the focal loss function truly enabled the CNNs models to be less biased towards the majority class than the cross-entropy did in the classification task of imbalanced dog red blood cell data.
\end{abstract}

\section{Introduction}
\label{Section:Introduction}
Human red blood cell morphology provides useful information for disease diagnosis. In the same vein, dog red blood cell morphology can give clues about a dog's health to veterinarians. There are five important features of red blood cell morphology classification: (i) shape, (ii) size, (iii) colour, (iv) inclusion, and (v) arrangement.  A morphology of red blood cells must be classified accurately in order for a veterinarian to apply an appropriate treatment~\citep{ref1}. Normally, a pathologist is needed to classify red blood cell morphologies by looking at the cells under a microscope. This standard method requires great expertise in manual classification. It is a very time-consuming qualitative and quantitative process that is prone to error~\citep{ref2}. Recently, computer vision and machine learning have been applied to human blood cell classification problem~\citep{ref2,ref3,ref4,ref5,ref6,ref14}. 

In the image classification field, conventional machine learning techniques require humans to manually extract useful features, i.e., converting raw data including image colour pixel values, image shape, and image texture into appropriate representations or feature vectors for a classifier to accurately classify the input image. Nonetheless, there is a method called representation learning that allows a machine to discover features from raw data automatically. 

Deep learning methods are representation-learning methods with multiple levels of representation layers that outperform conventional machine learning techniques in many kinds of tasks including image classification task~\citep{ref46, ref7}. Recently, several studies have employed Convolutional Neural Network (CNN) to tackle blood cell classification problem~\citep{ref11,ref12,ref15,ref20,ref44,ref45,ref21}. The results of those studies show that deep learning method is efficient in both blood cell feature extraction and classification. Most of those works focused on human blood cells. It has been known that the proportion of normal red blood cell samples found in the majority of people to abnormal red blood cell samples found in patients (the minority) is high. Moreover, the proportion of normal red blood cell samples to abnormal red blood cell samples is also high among patients themselves. To sum up, the rarity of every kind of abnormal red blood cell morphology causes the collected cell morphology data to be imbalanced~\citep{ref46}, which is a common problem in real-world datasets~\citep{ref22,ref23,ref25}. This leads deep learning to achieve high prediction accuracy for the majority class and poor prediction accuracy for the minority class~\citep{ref8}. In this classification task, though, detecting rare classes (minority class) is often more important~\citep{ref22}. To the best of our knowledge, based on the following pieces of literature~\citep{ref11,ref12,ref15,ref20,ref44,ref45,ref21}, there has been no research concerning class imbalance problem in classification of human blood cells.

This imbalanced data problem has been reported to reduce the performance of some classifiers~\citep{ref24}. Most existing learning algorithms did not perform well for minority class~\citep{ref25}. For over two decades, many class imbalance learning techniques have been developed~\citep{ref26}. Class imbalance learning can be divided into two main groups, namely, data-level and algorithm-level~\citep{ref27}. Common approaches in data-level group are resampling approaches~\citep{ref17,ref34,ref35,ref36,ref37,ref38,ref43} that aim to balance the training data. Re-sampling techniques can be further divided into three groups depending on the balancing class distribution method~\citep{ref9}:
\begin{itemize}
\item Over-sampling method: this method aims to increase minority class samples. Two widely used methods are randomly duplicating minority samples and generating synthetic minority samples.
\item Under-sampling method: this method discards some majority class samples in the dataset. The simplest method, random under-sampling, randomly removes majority class samples.
\item Hybrid method: this method is a combination of over-sampling and under-sampling methods.
\end{itemize}
A typical algorithm-level method is a cost-sensitive learning method~\citep{ref27,ref39,ref40,ref41,ref42}. It assigns a higher misclassification cost on the minority class~\citep{ref27}, making the classifier focus more on the minority class. Lately, there have been many approaches that implement new loss functions for deep imbalanced learning~\citep{ref10,ref13,ref28,ref29,ref30,ref31,ref32,ref33}. Unfortunately, there are some disadvantages to those approaches. In re-sampling approaches, over-sampling methods are computationally complex because of an increasing number of samples, while under-sampling methods may end up discarding important information~\citep{ref33}. A cost-sensitive learning method is more computationally efficient, but it is difficult to assign an appropriate cost for each class~\citep{ref9}. Most loss function implementations consider only binary classification problems. In this work, we addressed the problem at algorithm-level or loss function because of its lower computational complexity. In the past decade, numerous loss functions have been developed. We selected focal loss function in this study~\citep{ref10}. It was originally designed to handle highly-imbalanced classes in object detection tasks. This loss function has widely used in many kinds of tasks~\citep{ref13,ref47,ref48,ref49} but not in a medical classification task.

Our contributions are as follows:
\begin{itemize}
\item We tackled dog red blood cell morphology classification problem by using two well-known deep CNN architectures, e.g., Residual Network (ResNet)~\citep{ref18} and Densely Connected Convolutional Networks (DenseNet)~\citep{ref16}.
\item We applied a focal loss function for a multi-classification task to handle the class imbalance problem in deep CNNs and compared the performance of the focal loss function to that of the cross-entropy loss function, a conventional deep CNN loss function.
\item The proposed method was implemented in a well-designed framework for training deep CNN models and for hyperparameter tuning for class imbalance learning.
\end{itemize}

\section{Related Work}
\label{Section:RelatedWork}

\subsection{Blood Cell Classification Methods}
Many researchers have developed methods for automated red blood cell count and classification from images by using several image processing, machine learning, pattern recognition, and computer vision techniques. \cite{ref4} proposed an automated blood cell count system that uses image processing and pattern recognition techniques on histopathological images of red and white blood cells. They applied an image processing algorithm for cell segmentation and then differentiated between red blood cells and white blood cells by their size estimates. Their framework achieved 90~\% accuracy in red blood cell count. \cite{ref3} proposed a red blood cell classification method based on digital image processing. Several features related to shape, internal central pallor configuration of red blood cells, their circularity, and elongation were extracted and utilised in their proposed rule-based system. \cite{ref5} proposed a hybrid neural network classifier to separate between normal and abnormal red blood cells based on their shape and texture features. The accuracy of the proposed method in classifying normal and abnormal was 88.25~\%. Moreover, they attempted to classify four types of disease---Burr cell, Sickle cell, Horn cell, and Elliptocyte cell---and achieved 91~\% accuracy. \cite{ref6} compared three machine learning algorithms---k-Nearest Neighbour (kNN), Support Vector Machine (SVM), and Extreme Learning Machine (ELM)---in sickle cell anaemia detection. Firstly, they employed image processing techniques including grayscale image conversion, noise filtering, image enhancement, and morphological operations then fuzzy C means clustering to separate between normal and sickle cells. In that paper, they mainly extracted the following features: (i) geometrical feature, namely, metric value and elongation and (ii) statistical features such as mean, standard deviation, variance, skewness, and kurtosis. The results show that ELM classifier was superior to kNN and SVM.

Recently, deep CNN methods have been applied to tackle red blood cell classification task. \cite{ref12} presented an efficient deep contour aware segmentation approach based on a fully conventional network. They used CNN to extract features from each segmented cell for an ELM classifier. The accuracy on red blood cell classification was 94.71~\% and 98.68~\% on white blood cell classification. The accuracy on red blood cell segmentation was 98.12~\% and 98.16~\% on white blood cell segmentation. \cite{ref15} aimed to evaluate the performance of CNN in red blood cell morphology classification task. They employed some image augmentation techniques to increase the number of training images and then trained three DenseNet models with the same training set but with different random seed initialisers to evaluate the reproducibility and calculate the ensemble predictions. They concluded that DenseNet is suitable for red blood cell morphology classification. \cite{ref20} proposed an automatic detection and classification system for white blood cells from peripheral blood images. They first used different values of red and blue colours (R-B image) to separate white blood cells from red blood cells and then converted R-B image to binary image with a threshold value, making the nucleus more apparent, followed by a morphology operation to remove slight noise. Subsequently, a granularity feature (pairwise rotation invariant co-occurrence local binary pattern or PRICoLBP feature) was extracted to feed into the SVM to separate two classes---eosinophils and basophils---from one another. Consequently, CNN was applied to the other classes to extract features. The deep-learned features were inputted into a random forest classifier to classify the other three types of white blood cells: neutrophil, monocyte and lymphocyte. The proposed method achieved 92.8~\% classification accuracy. \cite{ref44} developed a deep learning model to handle a blood cell classification problem. They applied data augmentation to each white blood cell class, from 400 images to 3000 images, to enlarge the training data. Furthermore, they compared their proposed model with Na\"{\i}ve Bayes and SVM. The results show that their proposed model outperformed the traditional machine learning models.

\subsection{Class Imbalance Learning}
For the class imbalance problem, many research studies have mainly concentrated on two levels: data-level and algorithm-level~\citep{ref50}.

For the data level, the class imbalance problem was manipulated by either over-sampling the minority class or under-sampling the majority class. The general problem of the over-sampling method was that it only replicated the minority class samples randomly, leading to an over-fitting problem~\citep{ref17,ref50}. In order to solve the over-fitting problem, \cite{ref17} proposed a method called Synthetic Minority Over-sampling Technique (SMOTE) that over-sampling the minority class by generating new data points. Lately, \cite{ref36} introduced a Borderline-SMOTE algorithm, a modified form of SMOTE algorithm. This algorithm over-samples only the minority samples that are close to the class borderline. Another method called data augmentation has been used to over-sample the minority samples~\citep{ref45}. That paper presented a deep residual learning method for finer white blood cell classification. They applied data augmentation to balance the image samples among classes, e.g., flipped the images, randomly cropped the original images, added random noise to the original images, etc. However, increasing the training data led to a higher computational cost. On the other hand, under-sampling is preferred as \cite{ref51} demonstrated that under-sampling methods are more efficient than over-sampling methods~\citep{ref33,ref50}.

For algorithm level, an algorithm that emphasises the minority class was employed. A common approach for this kind of level is cost-sensitive learning. Most cost functions give equal importance to each class~\citep{ref50}. Therefore, a proper weight needs to be specified for each class. \cite{ref40} applied cost-sensitive learning with ELM. They presented weighted ELM for imbalance learning. Their experimental results show that the weighted ELM performed better than the unweighted ELM. \cite{ref27} aimed to tackle class imbalance problem by weighting the cost of each class in Deep Belief Network with an Evolutionary Algorithm. Their proposed method performed significantly better than the others on 58 benchmark datasets and a real-world dataset. Another approach is to modify the loss function in order to make the classifier more sensitive toward minority classes. \cite{ref10} proposed a focal loss function to tackle the extreme foreground-background class imbalance for dense object detection by modifying the conventional cross-entropy loss. They demonstrated that a one-stage detector RetinaNet using focal loss function yielded a better performance than those of faster RCNN models. \cite{ref13} proposed a semi-focal loss function, modified from a focal loss function, to handle erratic labelling problem in Mitosis Detection. Moreover, they proposed a new mitosis detection network called Cascaded Neural Network with Hard Example Mining and Semi-focal Loss. Their method achieved the best $F_{1}$-score at 0.68 in Tumor Proliferation Assessment Challenge 2016.

\section{Methodology}
\subsection{Cross-entropy loss}
Cross-entropy loss function is generally used in a deep learning classification model. It is defined as:
\begin{equation}
    CE(p,y) = CE(p_t) = -\log (p_t),
\label{eq1}
\end{equation}
where
\begin{equation}
p_{t} = \begin{cases}
    p,              & \text{if}\:$y = 1$;\\
    1-p             & {\text otherwise,}
\end{cases}
\end{equation}
$p$ is the prediction probability of the model, and $y$ is ground truth-label. 

\subsection{Focal loss}
\begin{figure}[!ht]
\centering
\includegraphics[width=0.6\columnwidth]{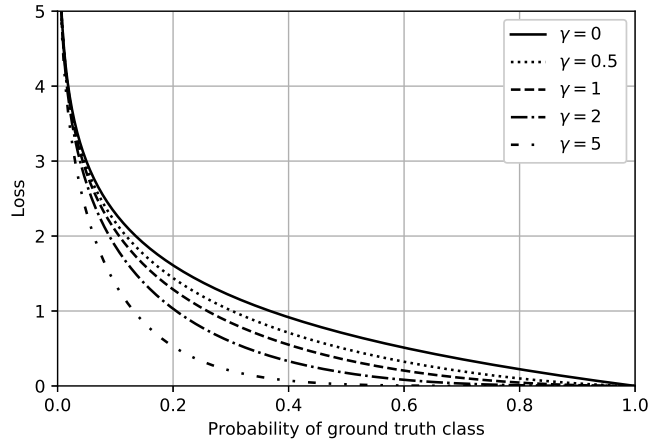}
\caption{The effect of a hyper-parameter on focal loss}
\label{Figure:FocalLoss}
\end{figure}

Focal loss is modified from cross-entropy loss by adding a modulating factor $(1-p_{t})^\gamma$ to the cross-entropy loss~\citep{ref10}. It is defined as
\begin{equation}
FL(p_{t}) = -(1-p_{t})^\gamma\log (p_{t}).
\label{eq2}
\end{equation}
In practice, a focal loss function uses an $\alpha$-balanced variant of focal loss:
\begin{equation}
FL(p_{t}) = -\alpha(1-p_{t})^\gamma\log (p_{t}).
\label{eq3}
\end{equation}
$\gamma$ is a focusing parameter that reshapes the loss function to down-weight easy samples and makes the model focus on hard samples. Hard samples are those samples that produce large errors; a model misclassifies the samples with a high probability. Figure~\ref{Figure:FocalLoss} shows the effect of a hyper-parameter on focal loss. When $\gamma=0$, the function behaves like $CE$---presented as a solid line in the figure. 

For cross-entropy loss, when a model classifies easy samples---samples in the majority class---to a correct class with $p_{t}$ $\geq$ 0.5, the loss value is low. Although the loss is low, when it is summed over a large number of easy samples, these loss values may overwhelm the rare class in an imbalance data scenario~\citep{ref10}. This can lead to a biased model towards the majority class. 

In focal loss, in the case that the ground truth label is 1 and a sample is correctly classified with a high probability, the value of $(1-p_{t})$ is small. When this term is raised to the power of $\gamma$, the value of the modulating factor gets smaller and causes the loss from cross-entropy to be smaller. In contrast, if the model misclassifies a sample with low probability, the modulating factor is large, close to 1; therefore, the loss from cross-entropy remains the same. 

\subsection{Models}\label{sec3}
In this research, we aimed to tackle the class imbalance problem in red blood cell morphology classification by using CNN in conjunction with a focal loss function. Many deep CNN architectures have been developed in the past decade. We selected ResNet~\citep{ref18} and DenseNet~\citep{ref16} models because both architectures have already been evaluated on ImageNet~\citep{ref19} consisting of 1,000 classes and gave outstanding results. Moreover, we applied a focal loss function to tackle the highly imbalanced data and compared the performances of both models between using a focal loss function and using a cross-entropy loss function, common in an image classification task.

\subsubsection{ResNet}
\begin{figure}[!ht]
\centerline{\includegraphics[width=0.5\columnwidth]{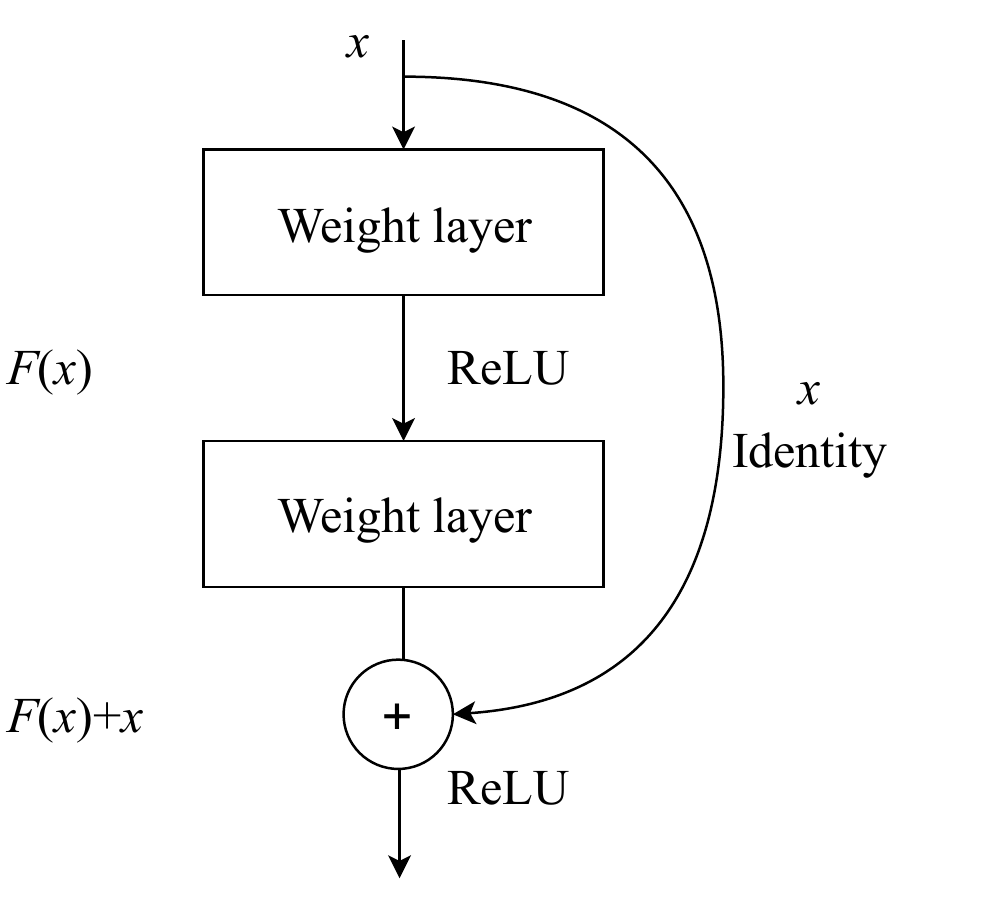}}
\caption{Residual block}
\label{Figure:ResidualBlock}
\end{figure}

\begin{figure}[!ht]
\centerline{\includegraphics[width=0.5\columnwidth]{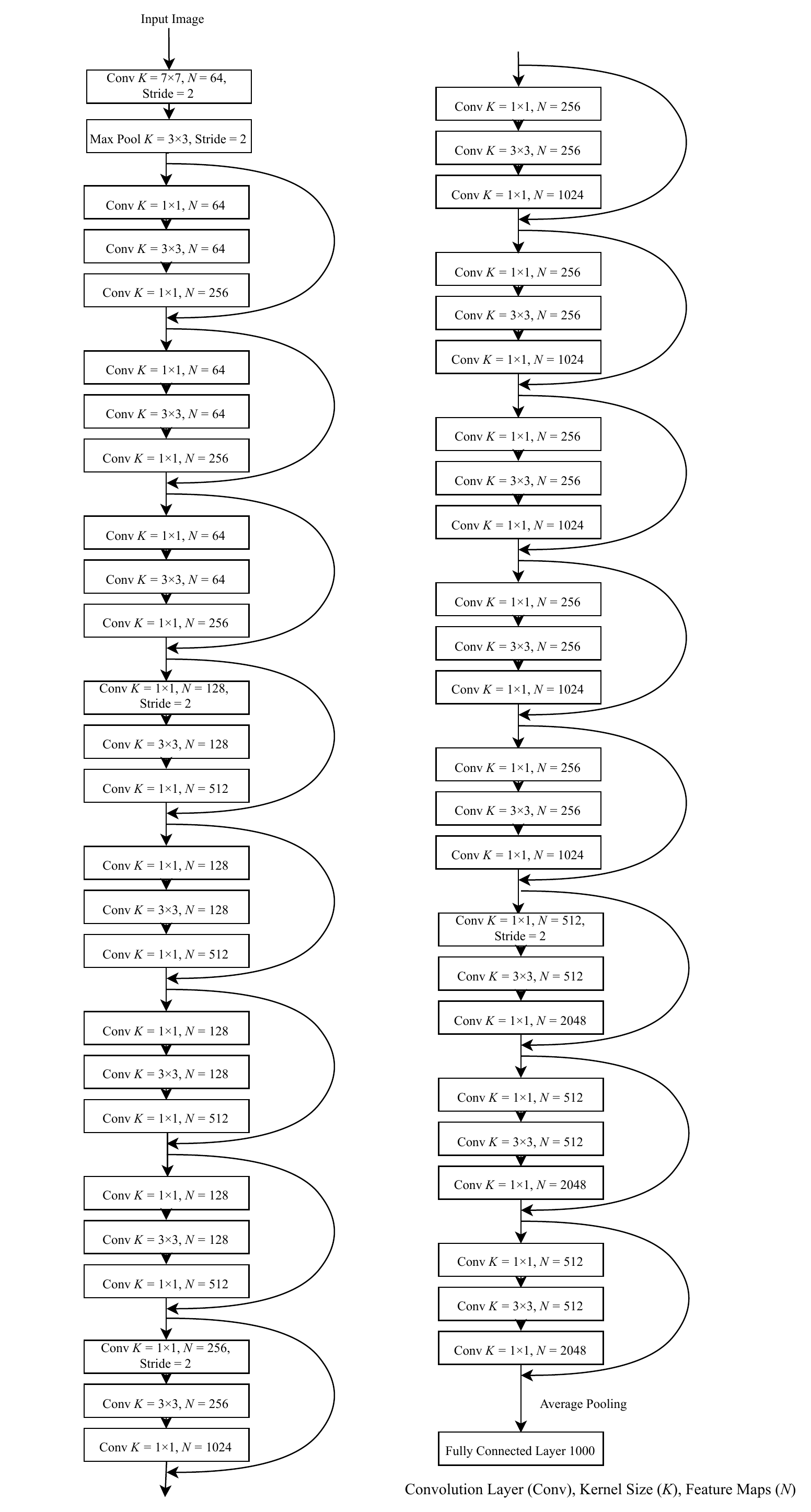}}
\caption{ResNet-50 architecture}
\label{Figure:ResNet50Architecture}
\end{figure}

It has been known that when a plain network is deep, the back-propagation gradient is small or vanishing, resulting in higher training and test errors~\citep{ref18,ref16,ref54,ref55,ref56}. To solve this vanishing gradient problem, \cite{ref18} proposed a ResNet model. ResNet architecture has a residual block that preserves the gradient. This is done by adding the input $x$ to the output after a few weight layers as shown in Figure~\ref{Figure:ResidualBlock}, enabling the model to pass useful knowledge from a previous layer to the next. Therefore, this enables the model to have less training error as the network is getting deeper. Furthermore, a ResNet model converges faster compared to a plain network. The experimental results show that ResNet-50, ResNet-101, and ResNet-152 performed better than VGG-16, GoogLeNet (Inception-v1), and PReLU-Net in top-1 error and top-5 error on ImageNet~\citep{ref18}. Here, we selected ResNet-50, that is 50 layers deep, for our experiment because it is the smallest model. The architecture is shown in Figure~\ref{Figure:ResNet50Architecture}. 

\subsubsection{DenseNet}
\begin{figure*}[ht]
\centering
    \begin{subfigure}{\textwidth}
      \centerline{\includegraphics[width=0.9\linewidth]{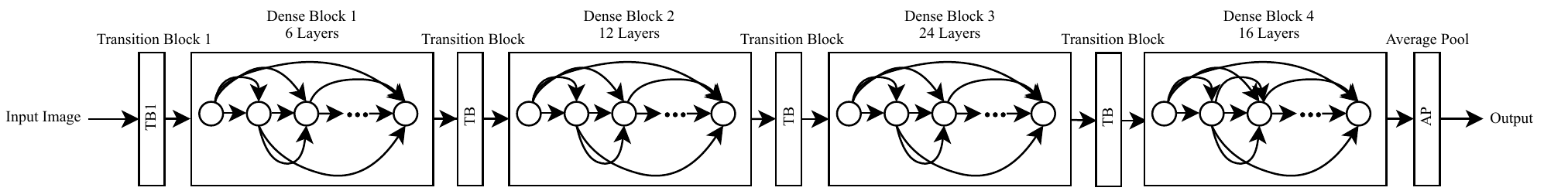}}  
      \caption{DenseNet with four dense blocks}
      \label{Figure:DenseNet121Architecturea}
    \end{subfigure}
    \newline
    \begin{subfigure}{\textwidth}
      \centerline{\includegraphics[width=0.8\linewidth]{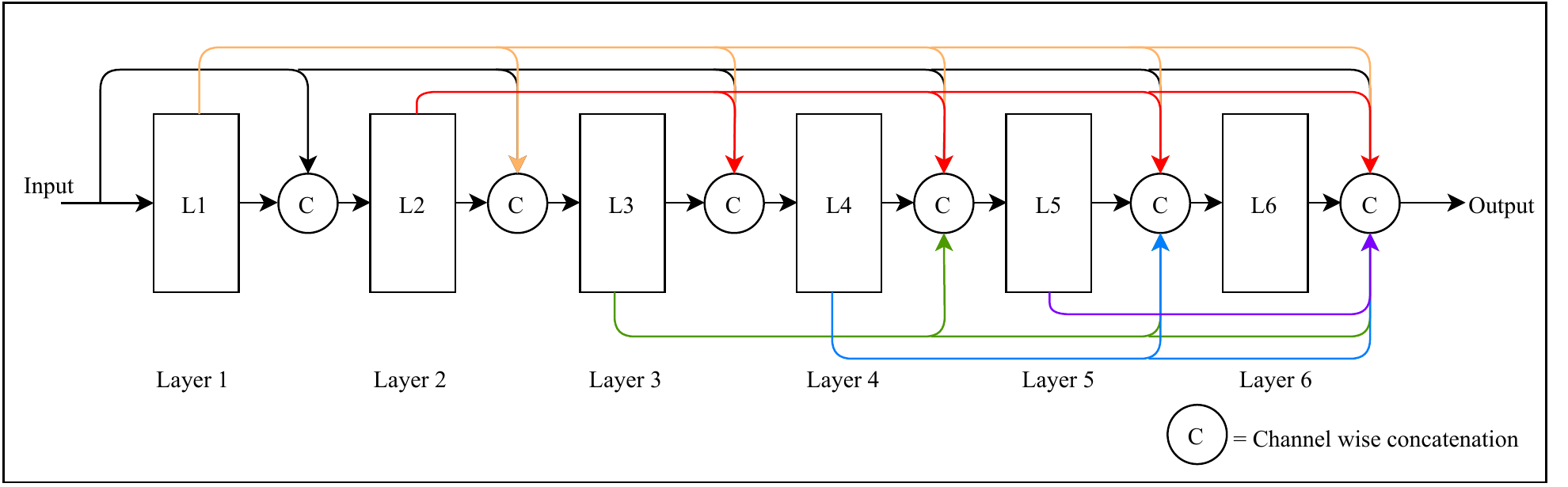}}  
      \caption{Dense Block 1 with 6 layers}
      \label{Figure:DenseNet121Architectureb}
    \end{subfigure}
    \newline
    \begin{subfigure}{\textwidth}
      \centerline{\includegraphics[width=0.8\linewidth]{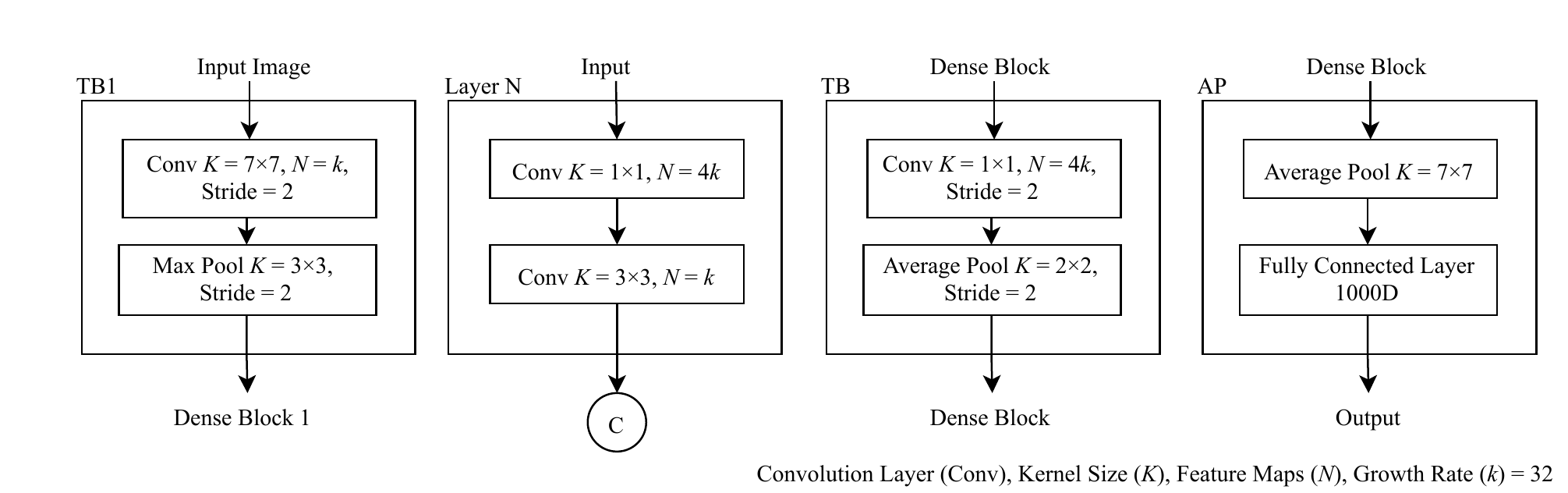}}  
      \caption{DenseNet components}
      \label{Figure:DenseNet121Architecturec}
    \end{subfigure}
\caption{DenseNet-121 architecture}
\label{Figure:DenseNet121Architecture}
\end{figure*}

DenseNet utilises dense connections, in which each layer receives extra inputs from all preceding layers, and also pass its own feature-map to all subsequent layers as shown in Figure~\ref{Figure:DenseNet121Architecture}. In ResNet, features are combined through summation before passing into a layer. On the other hand, DenseNet combines features by concatenating them to all subsequent layers. Each layer gains the collective knowledge of all other layers, resulting in a thinner and compact network. Moreover, DenseNet clearly achieved a higher accuracy with a smaller number of parameters than ResNet did~\citep{ref16}. We employed DenseNet-121 in this research because it is the smallest available architecture. The model is shown in Figure~\ref{Figure:DenseNet121Architecture}.

\section{Experimental Framework}
\subsection{Dataset}
\begin{figure}[!ht]
\centerline{\includegraphics[width=\columnwidth]{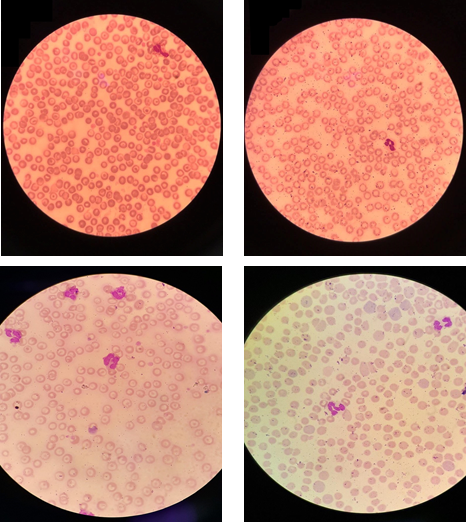}}
\caption{Dog blood smear images\label{Figure:SmearImage}}
\end{figure}

The 22 dog blood smear images were taken by a camera through a microscope at 100$\times$ magnification. The images were provided by the Veterinary Teaching Hospital, Kasetsart University, Hua Hin, Thailand. Examples of dog blood smear images are shown in Figure~\ref{Figure:SmearImage}.

\begin{figure*}[ht]
\centering
    \begin{subfigure}{0.45\textwidth}
      \centerline{\includegraphics[width=0.5\linewidth]{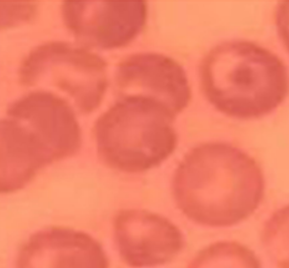}}  
      \caption{Conventional up-scale method}
      \label{Figure:CompareUpScaleMethoda}
    \end{subfigure}
    \begin{subfigure}{0.45\textwidth}
      \centerline{\includegraphics[width=0.5\linewidth]{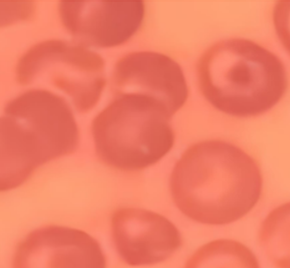}}  
      \caption{Waifu2$\times$}
      \label{Figure:CompareUpScaleMethodb}
    \end{subfigure}
\caption{Comparison between conventional up-scale method and Waifu2$\times$}
\label{Figure:CompareUpScaleMethod}
\end{figure*}

Since the default size of the input image for both ResNet-50 and DenseNet-121 was $224 \times 224$ pixels, we fed, one at a time, a single blood cell image with this size into the models. Therefore, we normalised the blood smear images to a size that each blood cell matches $224 \times 224$ pixels. Therefore, the input images were required to be enlarged by a scaling factor. A scaling factor was calculated by dividing 224 with the diameter of the red blood cell. If we simply applied the conventional up-scaling method, each image would not be clear and would be full of noise (as shown in Figure~\ref{Figure:CompareUpScaleMethoda}). Then, we employed an image super-resolution tool called ``Waifu2$\times$'' which was based on the CNNs~\citep{ref52} to up-scale all images. This technique enhanced the overall resolution of the output images as shown in Figure~\ref{Figure:CompareUpScaleMethodb}.

\begin{figure*}
\centering
\begin{subfigure}[t]{.3\textwidth}
  \centering
  \includegraphics[width=.8\linewidth]{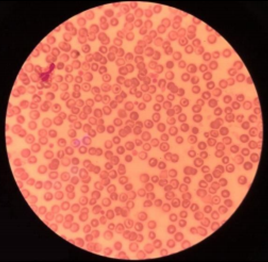}  
  \caption{Up-scaled blood smear image}
  \label{Figure:ImagePreprocessinga}
\end{subfigure}
\begin{subfigure}[t]{.3\textwidth}
  \centering
  \includegraphics[width=.8\linewidth]{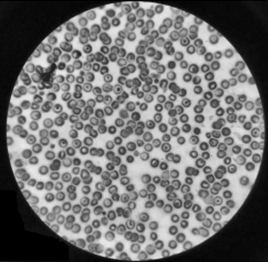}  
  \caption{Adaptive histogram equalisation algorithm was applied to (a)}
  \label{Figure:ImagePreprocessingb}
\end{subfigure}
\begin{subfigure}[t]{.3\textwidth}
  \centering
  \includegraphics[width=.8\linewidth]{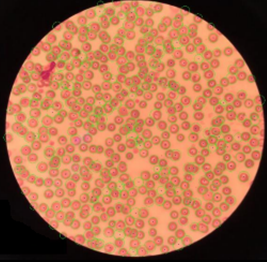}  
  \caption{Hough circle transform algorithm was applied to (b) for red blood cell detection}
  \label{Figure:ImagePreprocessingc}
\end{subfigure}
\par\bigskip
\begin{subfigure}{.3\textwidth}
  \centering
  \includegraphics[width=.5\linewidth]{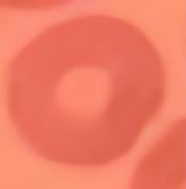}  
  \caption{Segmented cell images}
  \label{Figure:ImagePreprocessingd}
\end{subfigure}
\begin{subfigure}{.3\textwidth}
  \centering
  \includegraphics[width=.5\linewidth]{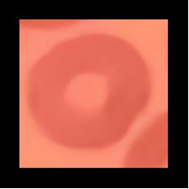}  
  \caption{Segmented cell with zero-padding}
  \label{Figure:ImagePreprocessinge}
\end{subfigure}
\caption{Cell segmentation process}
\label{Figure:ImagePreprocessing}
\end{figure*}

\begin{figure*}
\centering
\begin{subfigure}{1\textwidth}
  \centering
  \includegraphics[width=.7\linewidth]{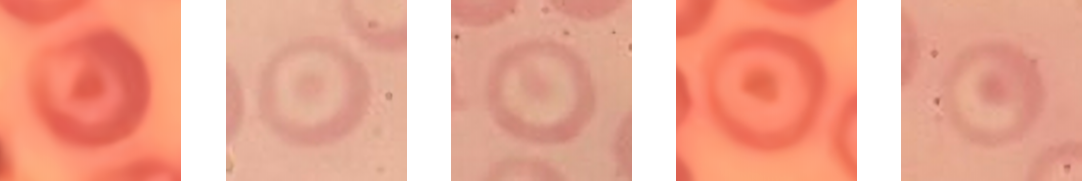}  
  \caption{Codocyte (Target Cell)}
  \label{Figure:ExampleOfRBCa}
\end{subfigure}
\newline
\begin{subfigure}{1\textwidth}
  \centering
  \includegraphics[width=.7\linewidth]{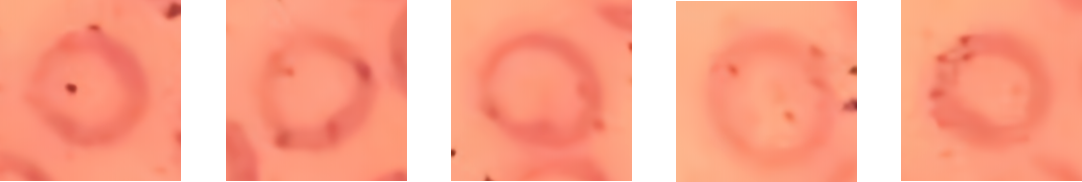}  
  \caption{Hypochromia}
  \label{Figure:ExampleOfRBCb}
\end{subfigure}
\newline
\begin{subfigure}{1\textwidth}
  \centering
  \includegraphics[width=.7\linewidth]{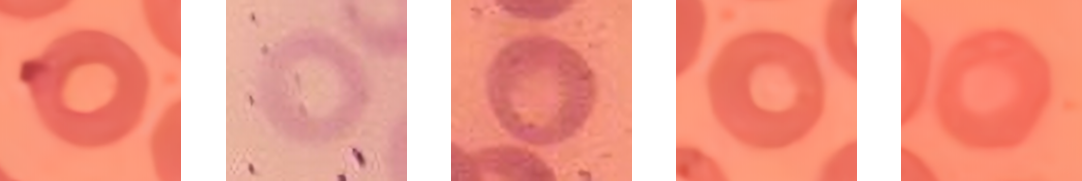}  
  \caption{Normal}
  \label{Figure:ExampleOfRBCc}
\end{subfigure}
\caption{Three classes of red blood cell morphology in this experiment}
\label{Figure:ExampleOfRBC}
\end{figure*}

After the image normalisation process, adaptive histogram equalisation algorithm was applied to increase the contrast between the background and the red blood cells as shown in Figure~\ref{Figure:ImagePreprocessingb}. Then, we employed a Hough circle transform algorithm to detect the red blood cells in a smear image as shown in Figure~\ref{Figure:ImagePreprocessingc}. Next, we segmented all red blood cells (as shown in Figure~\ref{Figure:ImagePreprocessingd}). These images were then transferred to a pathologist at Kasetsart University to label all the segmented cells. It should be noted that the segmented image size was close to $224 \times 224$. We added zero-padding to all segmented red blood cell images to make them $224 \times 224$ in size as shown in Figure~\ref{Figure:ImagePreprocessinge}. We describe the data pre-processing procedure in Algorithm~\ref{Algorithm:CellSegmentation}. All red blood cells were labelled and distinguished into three main groups by the pathologist. Our dataset contained 3,392 cell images divided into three classes: (i) 345 Codocyte cells (Target cells), (ii) 356 Hypochromia cells, and (iii) 2,691 Normal cells. Examples of red blood cell morphology images are shown in Figure~\ref{Figure:ExampleOfRBC}.

\begin{algorithm*}
\caption{Cell segmentation algorithm}\label{Algorithm:CellSegmentation}
\begin{algorithmic}[1]
    \ForAll{Smear images}
        \State UpscaleRatio $\gets \frac{\text{ExpectedCellDiameter}}{\text{CellDiameter}}$
        \State UpscaledImage $\gets$ Waifu2$\times$(SmearImage, UpscaleRatio)
        \State NormalisedImage $\gets$ Normalise(UpscaledImage) \Comment{Normalise to greyscale image}
        \State AdaptImage $\gets$ AdaptiveHistrogramEqualisation(NormalisedImage) \Comment{Increase contrast between background and cells}
        \State [CellsCoordinates, Radius] $\gets$ HoughCircleTransform(AdaptImage)
        \ForAll{\{CellsCoordinates, Radius\}}
            \State CellImage $\gets$ SegmentCell(CellCoordinates, Radius)
            \If{2 $\cdot$ Radius $\le$ ExpectedCellDiameter}
            \State PaddingSize $ \gets$ ExpectCellDiameter - (2 $\cdot$ Radius)
            \State PaddedImage $\gets$ ZeroPadding(CellImage, PaddingSize)
            \State CellImage $\gets$ PaddedImage
            \EndIf
            \State SaveImage(CellImage)
        \EndFor
        
    \EndFor

\end{algorithmic}
\end{algorithm*}

\subsection{Experiment Settings}\label{Section:Methodology:Settings}
\begin{figure*}[!ht]
\centerline{\includegraphics[width=\textwidth]{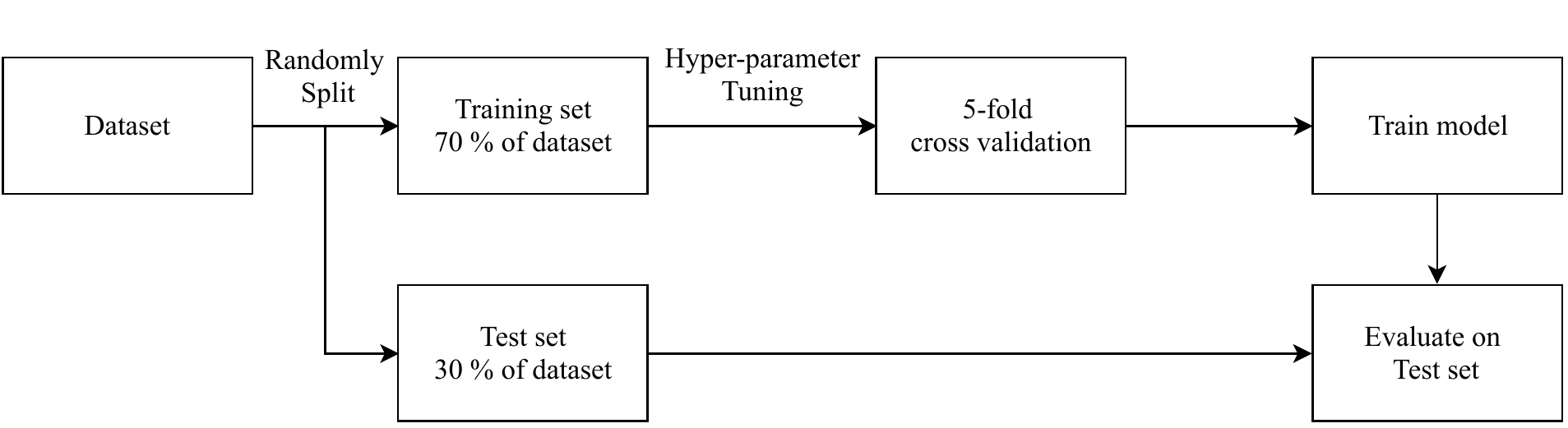}}
\caption{Process diagram}
\label{fig9}
\end{figure*}

We first randomly split 70~\% of our dataset as a training set and the remaining 30~\% as a test set. Since there are a number of hyper-parameters to be tuned, we utilised five-fold cross-validation to evaluate the settings of the models on the training set. In the focal loss function, there are two hyperparameters, namely, $\alpha$ and $\gamma$ as shown in~(\ref{eq3}). In this experiment, we simply set $\gamma=2$ because it worked the best in~\citep{ref10}. Therefore, we concern only on $\alpha$ that varies in the range of [0.25, 0.50, 0.75, 1.00, 1.25, 1.50, 1.75]. The cross-entropy function had no hyperparameter. However, there was still a model hyperparameter---the number of epochs---to be acquired. The number of epochs was determined based on the minimum value of the corresponding loss. Apart from this hyperparameter, we set the input batch size to 32 and the maximum number of epochs to 100. Adam method was utilised as the optimiser~\citep{ref53}, and the learning rate was set to 0.001 for both models. After we obtained the optimal set of parameters, we trained the models on the training set and evaluated the models on the test set as shown in Figure~\ref{fig9}.

We compared the models with focal loss function against the models with cross entropy function. In addition, we employed re-sampling techniques including over-sampling and under-sampling techniques in our task. For the under-sampling technique, we randomly discarded the majority class samples until the number of samples in the majority class was equal to the number of samples in the minority class. For the over-sampling technique, we employed an image augmentation technique by randomly rotating the image between $-15$ to $15$ degrees and randomly flipping the image left, right, up, and down to increase the number of samples in the minority class to be equal to the number of samples in the majority class.

It is known that accuracy is commonly used for model evaluation. However, under this imbalanced data scenario, a model could achieve high accuracy by predicting only the majority class. Therefore, we also reported the Area Under the Receiver Operating Characteristic Curve (AUROC) and $F_{1}$-score. These have been frequently used as evaluation measures for imbalanced data problems~\citep{ref9}.

\section{Results and Discussion}\label{Section:ResultsAndDiscussion}
We first investigated the effect of the focal loss function hyperparameter on the performances of the two models. It should be noted that the $\gamma$ hyperparameter of the focal loss function was fixed to 2 because it was a reported best value~\citep{ref10}. We varied $\alpha$ from 0.25 to 1.75 with 0.25 step size and reported accuracy, $F_{1}$-scores and AUROC, shown in Table~\ref{Table:ExperimentalResults:All}. The $F_1$-Score achieved by both models with focal loss function slightly increased as $\alpha$ increased, until it reached a peak at $\alpha$ = 1.5 for DenseNet-121 and $\alpha$ = 1.0 for ResNet-50. DenseNet-121 with focal loss function outperformed the models with cross-entropy loss function in 4/7 cases, while Resnet-50 with focal loss function did so in 2/7 cases. We further compared the performances of ResNet-50 and DenseNet-121 with focal loss function to the two models with conventional cross-entropy loss function in combination with re-sampling  techniques. All functions were run with their optimal hyperparameters. Shown in Table~\ref{Table:ExperimentalResults:All}, DenseNet-121 model with focal loss function achieved the best performance with 0.92 $F_{1}$-score. It can also be seen that employing focal loss function in both models clearly improved their performances compared to those of both models with cross entropy loss function. Furthermore, the deep learning models with either kind of loss function outperformed the models with cross entropy loss function in combination with the under-sampling technique, as expected because of the smaller number training samples. It can be noticed that the use of cross-entropy loss function in combination with the over-sampling technique was able to improve the overall performances of the models to be in line with the performances of both models with focal loss function. DenseNet-121 achieved a 0.92 $F_{1}$-score with focal loss function and 0.91 $F_{1}$-score with cross-entropy loss function in combination with the over-sampling  technique. Hence, employing the over-sampling technique (image augmentation) was truly able to improve the performance of models with cross entropy functions. This is because over-sampling increased the number of samples in the training phase and successful deep models depended on a large number of samples~\citep{PasupaSunhem:2016}. It should be noted, though, that performing the over-sampling technique led to a higher computational cost. Overall, incorporating a focal loss function into the models clearly improved their performance.

\begin{table*}[!ht]
\caption{Accuracy, $F_{1}$-score, and AUROC achieved by the models with the optimal epoch for each hyperparameter $\alpha$ in the focal loss and cross-entropy loss functions. The best performances are expressed in bold text.}
\label{Table:ExperimentalResults:All}
\centering
\begin{tabular}{lccccccr}
\toprule
{Model architecture}&{Method}&\multicolumn{2}{@{}c@{}}{{Hyperparameters}} & {Optimal epoch} & {Accuracy} &{$F_{1}$-score} &{AUROC} \\
\cmidrule{3-4}
 &   & {$\alpha$}  & {$\gamma$}  & & ( \% ) & & \\
\midrule
\multirow{10}{*}{DenseNet-121} & \multirow{7}{*}{Focal loss} & 0.25  & {\multirow{7}{*}{2}}  & 89 &91.94 & 0.87 & 0.98 \\
                                &                                   & 0.50  &   & 68 &82.81 & 0.72& 0.96 \\
                                &                                   & 0.75  &   & 98 &88.80 & 0.78 & 0.96 \\
                                &                                   & 1.00  &   & 76 &92.24 & 0.85 & 0.98 \\
                                &                                   & 1.25  &   & 76 &92.93 & 0.86 & 0.98 \\
                                &                                   & {1.50}  &   & {88} &\textbf{95.60}& \textbf{0.92}& \textbf{0.99} \\
                                &                                   & 1.75  &   & 64 &83.01 & 0.55 & 0.95 \\
\cline{2-8}
                                & Cross-entropy loss & - & - & 75 & 91.30 & 0.84 & 0.95 \\
                                & Under-sampling & - & - & 93 & 88.70 & 0.80 & 0.97\\
                                & Over-sampling & - & - & 95 & 95.28 & 0.91 & 0.99\\
\cline{1-8}
\multirow{10}{*}{ResNet-50} & \multirow{7}{*}{Focal loss} & 0.25  & {\multirow{7}{*}{2}}  & 34 &83.50 & 0.61 & 0.93 \\
                                &                                   & 0.50  &   & 83 &86.44& 0.71 & 0.96 \\
                                &                                   & 0.75  &   & 36 &89.49& 0.81 & 0.94 \\
                                &                                   & {1.00}  &   & {90} &92.80& \textbf{0.87} & 0.97 \\
                                &                                   & 1.25  &   & 79 &92.04& 0.85 & 0.97 \\
                                &                                   & 1.50  &   & 93 &89.19& 0.83 & 0.97 \\
                                &                                   & 1.75  &   & 73 &85.17& 0.78 & 0.93 \\
\cline{2-8}
                                & Cross-entropy loss & - & - & 57 & 91.70 & 0.85 & 0.96 \\
                                & Under-sampling & - & - & 68 & 84.68 & 0.78 & 0.94\\
                                & Over-sampling & - & - & 75 & \textbf{92.93} & \textbf{0.87} & \textbf{0.98}\\
\bottomrule
\end{tabular}

\end{table*}

\begin{figure*}
\begin{subfigure}{.5\textwidth}
  \centering
  \includegraphics[width=.8\linewidth]{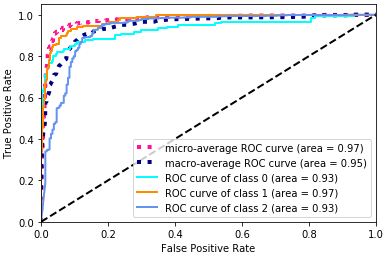}  
  \caption{DenseNet-121 with cross-entropy loss function}
  \label{fig11a}
\end{subfigure}
\begin{subfigure}{.5\textwidth}
  \centering
  \includegraphics[width=.8\linewidth]{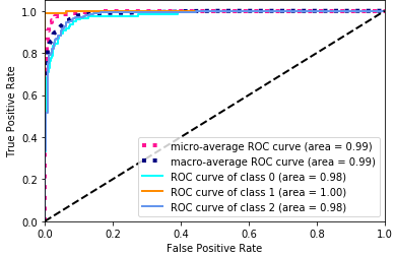}  
  \caption{DenseNet-121 with focal loss function($\alpha$=1.5 $\gamma$=2.0)}
  \label{fig11b}
\end{subfigure}
\newline
\begin{subfigure}{.5\textwidth}
  \centering
  \includegraphics[width=.8\linewidth]{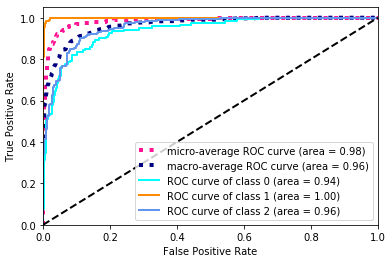}  
  \caption{ResNet-50 with cross-entropy loss function}
  \label{fig11c}
\end{subfigure}
\begin{subfigure}{.5\textwidth}
  \centering
  \includegraphics[width=.8\linewidth]{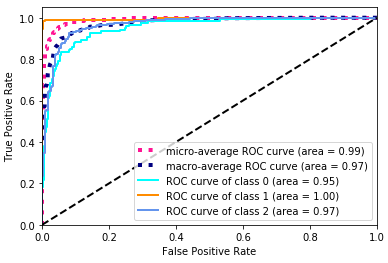}  
  \caption{ResNet-50 with focal loss function($\alpha$=1.0 $\gamma$=2.0)}
  \label{fig11d}
\end{subfigure}
\caption{Comparison of AUROC achieved by models with cross-entropy loss function and models with focal loss function.
 (class 0 is Codocyte (Target Cell), class 1 is Hypochromia, and class 2 is Normal)}
\label{fig11}
\end{figure*}

Next, we discuss the AUROC and confusion matrix of the best model. Figure~\ref{fig11} shows that both models with focal loss function achieved higher AUROC (near 1) for all classes than those achieved by models with cross-entropy loss function. This indicates that models with focal loss function were able to separate out each class very well in a class imbalance scenario.

\begin{figure*}
\begin{subfigure}{.5\textwidth}
  \centering
  \includegraphics[width=\linewidth]{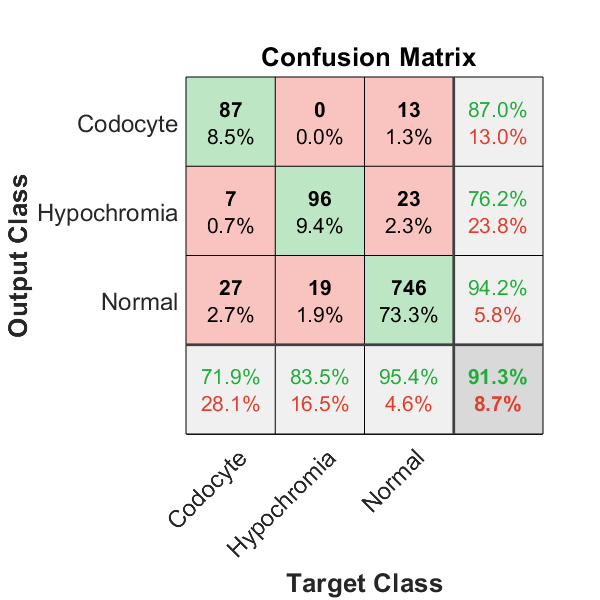}  
  \caption{DenseNet-121 with cross-entropy loss function}
  \label{fig12a}
\end{subfigure}
\begin{subfigure}{.5\textwidth}
  \centering
  \includegraphics[width=\linewidth]{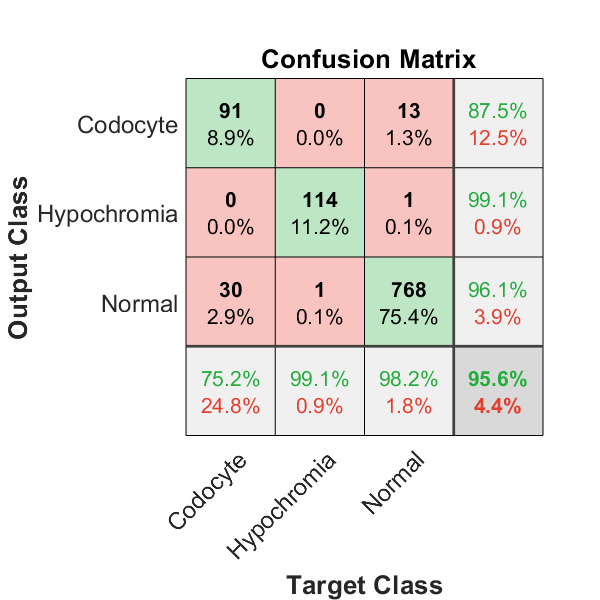}  
  \caption{DenseNet-121 with focal loss function ($\alpha$=1.5 $\gamma$=2.0)}
  \label{fig12b}
\end{subfigure}
\newline
\begin{subfigure}{.5\textwidth}
  \centering
  \includegraphics[width=\linewidth]{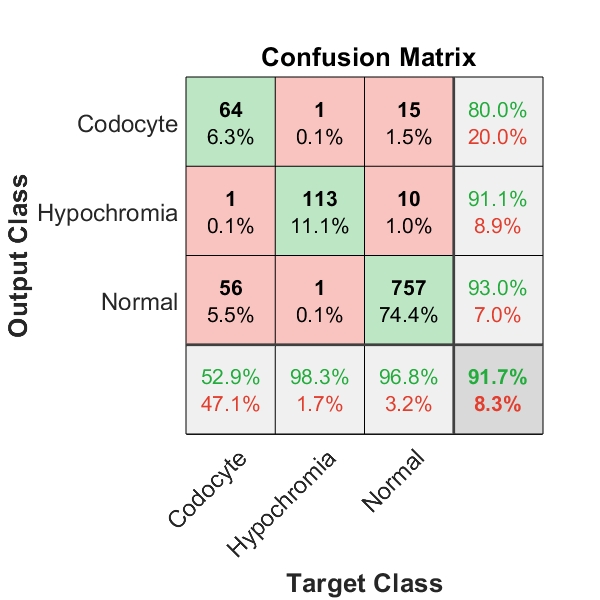}  
  \caption{ResNet-50 with cross-entropy function}
  \label{fig12c}
\end{subfigure}
\begin{subfigure}{.5\textwidth}
  \centering
  \includegraphics[width=\linewidth]{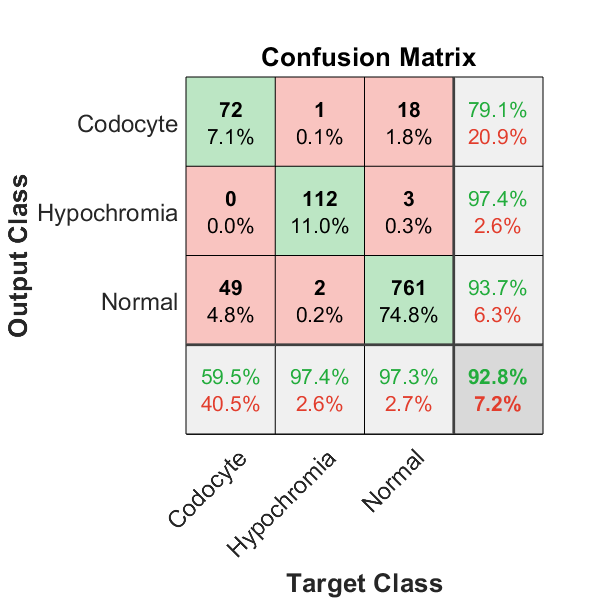}  
  \caption{ResNet-50 with focal loss function ($\alpha$=1.0 $\gamma$=2.0)}
  \label{fig12d}
\end{subfigure}
\caption{Comparison of the confusion matrix of both models with either cross-entropy loss function or focal loss function.}
\label{fig12}
\end{figure*}

In addition, we analysed the confusion matrix of models with either kind of loss function. The results are shown in Figure~\ref{fig12}. It should be noted that two minority classes were Codocyte that made up only 10.17~\% of the whole samples and Hypochromia that made up only 10.50~\% of the whole samples. ResNet-50 model with cross-entropy loss function achieved 52.9~\% accuracy in classifying Codocyte and 98.3~\% in classifying Hypochromia. The model did not perform well on Codocyte because it misclassified Codocyte to be Normal class for 5.5~\% of the total number of samples. This implies that the model was biased towards the majority class. However, when focal loss function was applied, the model did better in classifying Codocyte class and Normal class. This improved the model accuracy by 6.6~\%. Similarly, DenseNet-121 with cross-entropy loss function has an issue with classifying between the two minority classes and the majority class. DenseNet-121 achieved only 71.9~\% accuracy in classifying Codocyte and 83.5~\% accuracy in classifying Hypochromia. This is the common effect of the class imbalance problem. Thus, we incorporated the focal loss function into our model. Then, it was able to differentiate both minority classes more evidently and with high improvement. The accuracy in classifying Codocyte increased by 3.3~\% and in classifying Hypochromia increased by 15.6~\%. 

\begin{figure*}[!ht]
    \begin{subfigure}[t]{.5\linewidth}
    \centering
    \includegraphics[width=\columnwidth]{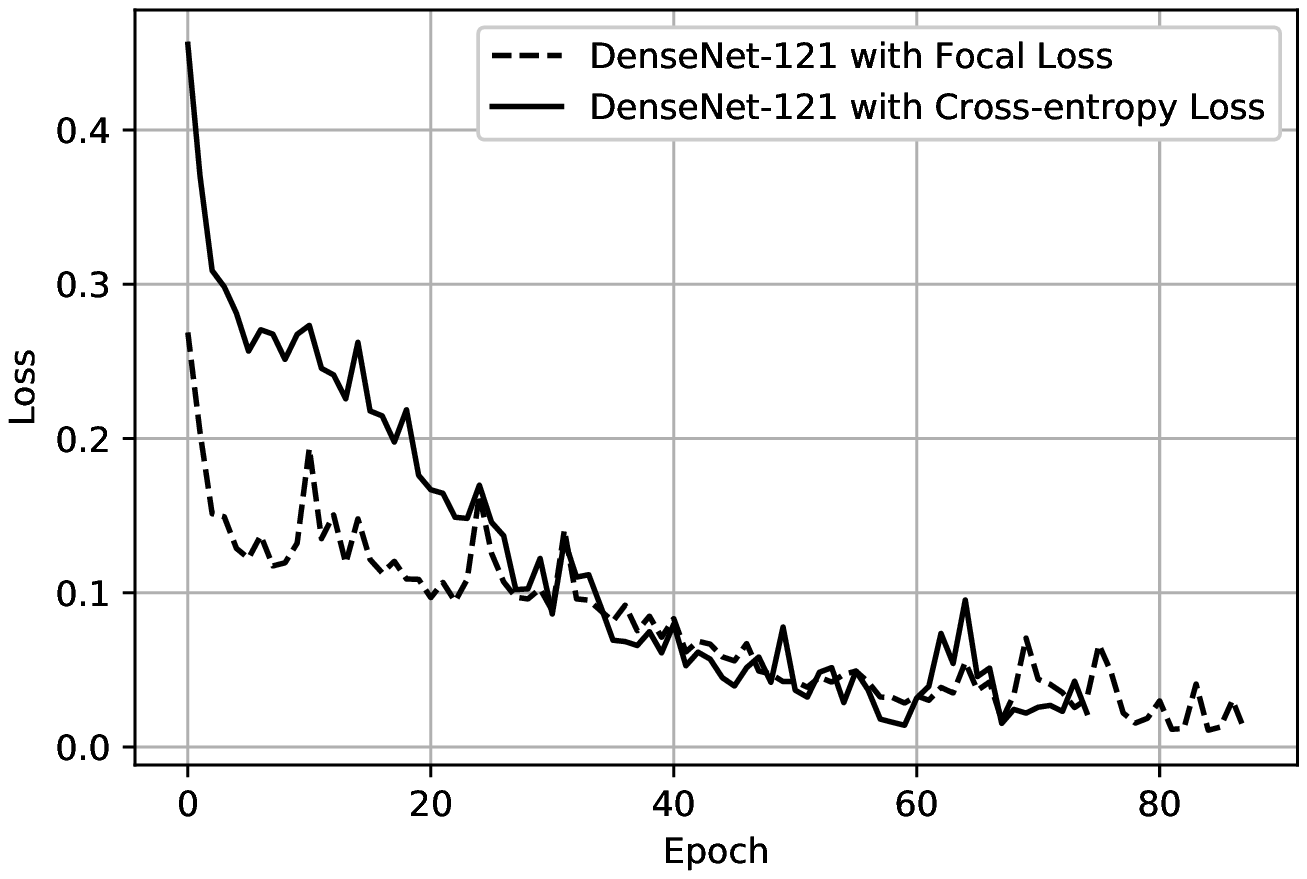}
    \caption{Comparison of training losses produced by DenseNet-121 with either the focal loss function or the cross-entropy loss function}
    \label{fig13a}
    \end{subfigure}
    \begin{subfigure}[t]{.5\linewidth}
    \centering
    \includegraphics[width=\columnwidth]{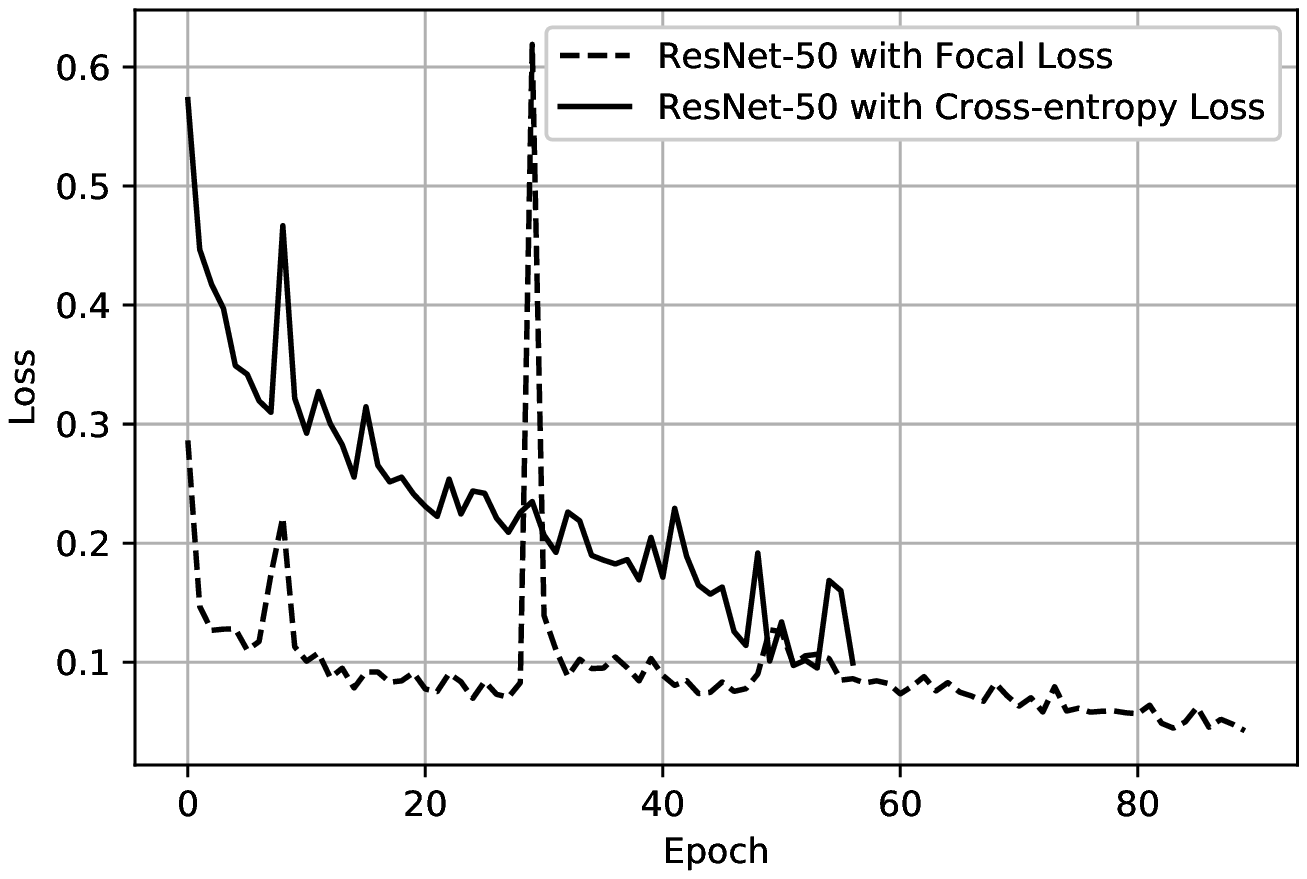}
    \caption{Comparison of training losses produced by ResNet-50 with either the focal loss function or the cross-entropy loss function}
    \label{fig13b}
    \end{subfigure}
\caption{Comparison of training losses produced by both models with either the focal loss function or the cross-entropy loss function}
\label{fig13}
\end{figure*}

Furthermore, we compared the training losses of both models with focal loss and cross-entropy loss functions in Figure~\ref{fig13}. Focal loss function is a generalised version of cross-entropy loss function; therefore, they can be plotted along the same co-ordinate axis. It can be clearly seen that the focal loss function enabled the training loss to converge to zero faster than the cross-entropy loss function could.

Furthermore, we examined the performances of the model with either the focal loss function or the cross-entropy loss function in dealing with class imbalance at different levels, shown in Table~\ref{Table:ClassDistribution}. The class distribution ratio of the original dataset was 10:10:80. To simulate a higher level of imbalance, we randomly removed samples in the minority classes and fixed the number of samples in the majority class. Conversely, we randomly removed samples in the majority class while fixing the number of samples in the minority classes to lower the level of imbalance. We calculated the imbalance ratio $(\rho)$~\citep{ref57} based on the distributions we obtained in Table~\ref{Table:ClassDistribution} by 
\begin{equation}
    \rho = \frac{\max_{i}(\left|{C_i}\right|)}{\min_{i}(\left|{C_i}\right|)},
\end{equation}
where ${C_i}$ is the set of samples in class $i$, $\max_{i}(\left|{C_i}\right|)$ and $\min_{i}(\left|{C_i}\right|)$ are the maximum and the minimum number of samples in all classes. Then, we randomly split 70~\% of each manipulated dataset to be a training set and the remaining 30~\% to be a test set. The model was then trained with the same settings described in Section~\ref{Section:Methodology:Settings}.

\begin{table*}[h]
\caption{Class distributions where $|C_i|$ is the number of samples in $C_i:i \in \{0,1,2\}$ ($C_0$ is Codocyte; $C_1$ is Hypochromia; and $C_2$ is Normal case).}
\label{Table:ClassDistribution}
\centering
\begin{tabular}{lccccccr}
\toprule
\multicolumn{3}{c}{{Class Distributions}} &{\multirow{2}{*}{$\rho$}}& \multirow{2}{*}{$|C_0|$} &\multirow{2}{*}{$|C_1|$} &\multirow{2}{*}{$|C_2|$} &\multirow{2}{*}{$\sum_{i=0}^{3}{|C_i|}$} \\
\cmidrule{1-3}
${C_0}$&${C_1}$&${C_2}$  & &  &  &  &  \\
\midrule
25.0 & 25.0 & 50.0 & 2.03 & 345 & 356 & 701 & 1402 \\
20.0 & 20.0 & 60.0 & 3.05 & 345 & 356 & 1052 & 1753 \\
15.0 & 15.0 & 70.0 & 4.74 & 345 & 356 & 1636 & 2337 \\
10.0 & 10.0 & 80.0 & 7.80 & 345 & 356 & 2691 & 3392 \\
 7.5  &  7.5  & 85.0 & 11.35 & 237 & 237 & 2691 & 3165 \\
 5.0  &  5.0  & 90.0 & 17.94 & 150 & 150 & 2691 & 2991 \\
 2.5  &  2.5  & 95.0 & 37.90&  71 &  71 & 2691 & 2833 \\
\bottomrule

\end{tabular}
\end{table*}

Employing the focal loss function on DenseNet-121 and ResNet-50 yielded better performances than using cross-entropy loss function for all $\rho$, resulting in higher $F_{1}$-scores as shown in Table~\ref{Table:ExperimentalResults:VaryDist}. Nevertheless, there were some cases that the models achieved a higher $F_{1}$-score but a lower accuracy and AUROC, e.g., DenseNet-121 with $\rho = 37.90$ and ResNet-50 with $\rho = 11.35$. Owing to these cases, it can be interpreted that the focal loss function made the deep CNN models learn to classify each class equally without biased towards the majority class. In contrast, the cross-entropy loss function provided a higher accuracy and AUROC score but lower $F_{1}$-score, resulting in a biased model towards the majority class.

\begin{table*}[!ht]
\caption{Comparison of the performances of DenseNet-121 and ResNet-50 with different $\rho$ with either focal loss function or cross-entropy loss function. Both models were run with their optimal hyperparameters. The best performance for each model is expressed as bold text.}
\label{Table:ExperimentalResults:VaryDist}
\resizebox{\textwidth}{!}{
\begin{tabular}{ccccccccccccc}
\toprule
\multirow{4}{*}{{$\rho$}}& \multicolumn{12}{c}{Model Architecture}\\
\cmidrule{2-13}
    & \multicolumn{6}{c}{DenseNet-121}   & \multicolumn{6}{c}{ResNet-50}\\
    \cmidrule(lr){2-7}
    \cmidrule(lr){8-13}
    & \multicolumn{3}{c}{Focal Loss}    & \multicolumn{3}{c}{Cross-entropy Loss} & \multicolumn{3}{c}{Focal Loss}    & \multicolumn{3}{c}{Cross-entropy Loss}\\
    \cmidrule(lr){2-4}
    \cmidrule(lr){5-7}
    \cmidrule(lr){8-10}
    \cmidrule(lr){11-13}
    & Accuracy & $F_{1}$-score & AUROC & Accuracy & $F_{1}$-score & AUROC & Accuracy & $F_{1}$-score & AUROC & Accuracy & $F_{1}$-score & AUROC \\
\midrule
2.03    & \textbf{89.79} & \textbf{0.90} & \textbf{0.97} & 86.22 & 0.84 & \textbf{0.97} & \textbf{88.36} & \textbf{0.88} & \textbf{0.97} & 88.36 & 0.87 & 0.97 \\
3.05    & \textbf{91.63} & \textbf{0.90} & \textbf{0.98} & 89.92 & 0.88 & 0.97 & \textbf{89.16} & \textbf{0.88} & \textbf{0.97} & 83.27 & 0.81 & 0.94 \\
4.74    & \textbf{87.61} & \textbf{0.84} & \textbf{0.95} & 87.32 & 0.83 & \textbf{0.95} & \textbf{90.60} & \textbf{0.86} & \textbf{0.97} & 90.03 & 0.84 & 0.95\\
7.80    & \textbf{95.60} & \textbf{0.92} & \textbf{0.99} & 91.30 & 0.84 & 0.95 & \textbf{92.8} & \textbf{0.87} & \textbf{0.97} & 91.7 & 0.85 & 0.96 \\
11.35   & \textbf{96.32} & \textbf{0.91} & \textbf{0.99} & 92.00 & 0.79 & 0.97 & 92.11 & \textbf{0.81} & \textbf{0.95} & \textbf{93.37} & 0.80 & 0.94 \\
17.94   & \textbf{96.10} & \textbf{0.85} & \textbf{0.98} & 93.99 & 0.78 & 0.97 & \textbf{94.32} & \textbf{0.79} & \textbf{0.94} & 92.43 & 0.78 & \textbf{0.94} \\
37.90   & 94.00 & \textbf{0.59} & 0.92 & \textbf{96.12} & 0.57 & \textbf{0.94} & \textbf{94.82} & \textbf{0.56} & \textbf{0.92} & \textbf{94.82} & 0.51 & 0.88 \\
\bottomrule

\end{tabular}}
\end{table*}

\begin{figure}[!ht]
\centering
\includegraphics[width=0.7\columnwidth]{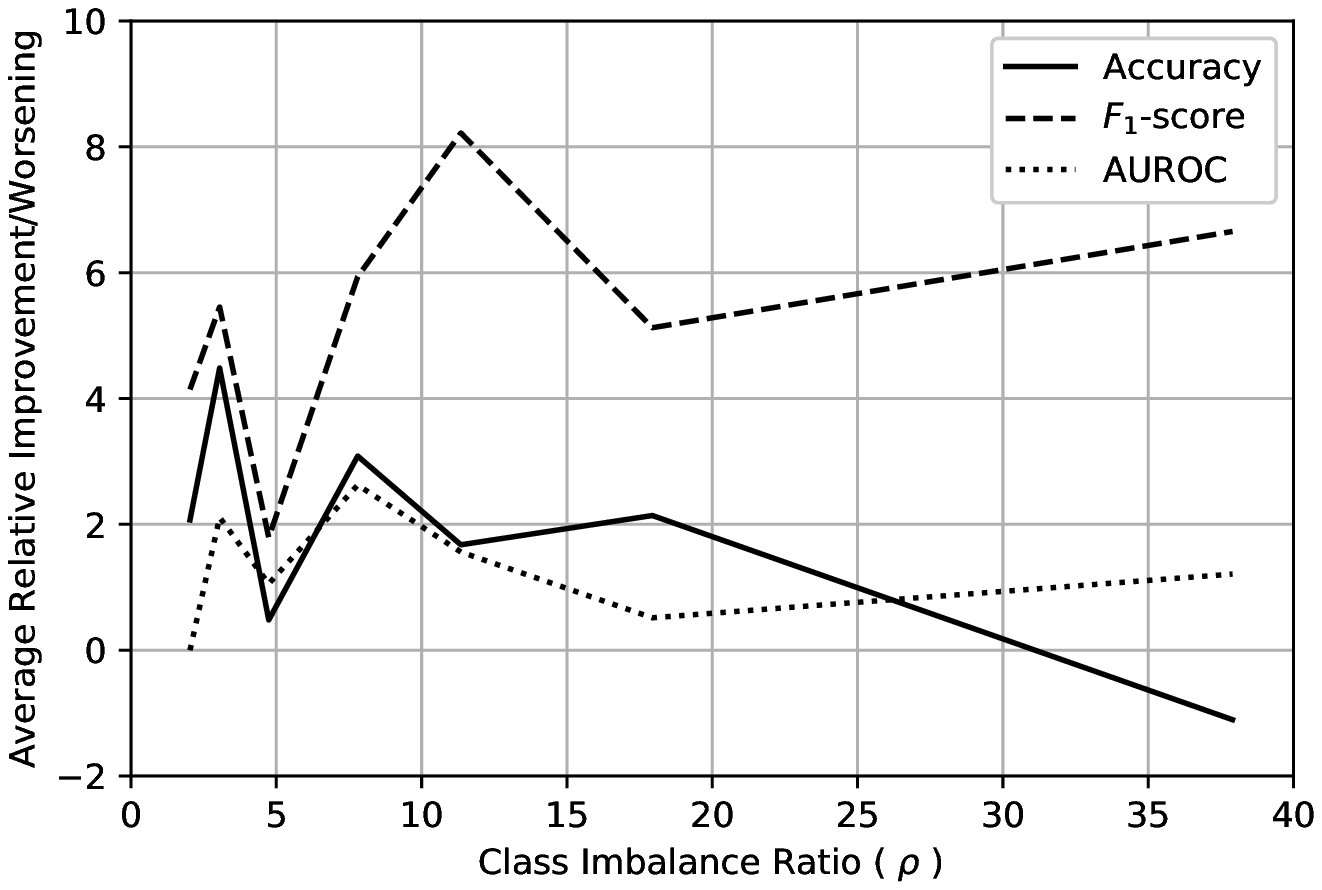}
\caption{Average Relative Improvement/Worsening of Accuracy, $F_{1}$-score and AUROC of CNN models with focal loss function against CNN models with cross-entropy loss function}
\label{Figure:Results:RelativeImprovement}
\end{figure}

We calculated and plotted the relative improvement and worsening in classification performance of DenseNet-121 and ResNet-50 achieved by incorporation of either focal loss function or cross-entropy loss function by averaging the relative improvement across both models in Figure~\ref{Figure:Results:RelativeImprovement}. The models with the focal loss function showed a higher improvement in $F_{1}$-score than in accuracy and AUROC. In addition, the relative improvement of $F_{1}$-score tended to increase as $\rho$ got higher. In contrast, the relative improvement in accuracy of the models with focal loss function tended to decrease because the biased model with cross-entropy loss function gained a higher accuracy but a lower $F_{1}$-score as $\rho$ got higher, i.e., the biased model failed to correctly classify the minority class.

\section{Conclusion}\label{sec6}
Class imbalance is commonly encountered in real-world data, particularly in medical image data. It makes a classification model easily biased towards the majority class. In addition, an inaccurate prediction may lead to false disease diagnosis. Hence, a way to deal with imbalanced data properly is very important. In this work, we proposed a method to solve class imbalance issue by incorporating a focal loss function into deep CNNs. Furthermore, we proposed a well-designed framework for class imbalance learning. We demonstrated that CNN models using a focal loss function achieved a higher $F_{1}$-score than the CNNs using a cross-entropy loss function. In addition, DenseNet-121 performed better than ResNet-50 and used fewer model parameters when using a focal loss function.

\begin{acknowledgement}
We would like to thank the Veterinary Teaching Hospital, Kasetsart University, Hua Hin, Thailand for providing us with stained glass slides of peripheral blood smear and for labelling the dataset.
\end{acknowledgement}

%
%

\bibliographystyle{myspbasic}      


\begin{thebibliography}{58}
	\providecommand{\natexlab}[1]{#1}
	\providecommand{\url}[1]{{#1}}
	\providecommand{\urlprefix}{URL }
	\expandafter\ifx\csname urlstyle\endcsname\relax
	\providecommand{\doi}[1]{DOI~\discretionary{}{}{}#1}\else
	\providecommand{\doi}{DOI~\discretionary{}{}{}\begingroup
		\urlstyle{rm}\Url}\fi
	\providecommand{\eprint}[2][]{\url{#2}}
	
	\bibitem[{Abulnaga and Rubin(2018)}]{ref49}
	Abulnaga SM, Rubin J (2018) Ischemic stroke lesion segmentation in {CT}
	perfusion scans using pyramid pooling and focal loss. In: Proceedings of the
	International MICCAI Brainlesion Workshop (BrainLes 2018), Granada, Spain,
	Springer, pp 352--363, \doi{10.1007/978-3-030-11723-8_36}
	
	\bibitem[{Baloch  \emph{et~al.}(2019)Baloch, Kumar, Haresh, Rehman, and
		Syed}]{ref33}
	Baloch BK, Kumar S, Haresh S, Rehman A, Syed T (2019) Focused anchors loss:
	cost-sensitive learning of discriminative features for imbalanced
	classification. In: Proceedings of the 11th Asian Conference on Machine
	Learning (ACML 2019), Nagoya, Japan
	
	\bibitem[{Buda  \emph{et~al.}(2018)Buda, Maki, and Mazurowski}]{ref57}
	Buda M, Maki A, Mazurowski MA (2018) A systematic study of the class imbalance
	problem in convolutional neural networks. Neural Networks 106:249--259,
	\doi{10.1016/j.neunet.2018.07.011}
	
	\bibitem[{Castro and Braga(2013)}]{ref41}
	Castro CL, Braga AP (2013) Novel cost-sensitive approach to improve the
	multilayer perceptron performance on imbalanced data. IEEE Transactions on
	Neural Networks and Learning Systems 24(6):888--899,
	\doi{10.1109/TNNLS.2013.2246188}
	
	\bibitem[{Chawla  \emph{et~al.}(2002)Chawla, Bowyer, Hall, and
		Kegelmeyer}]{ref17}
	Chawla NV, Bowyer KW, Hall LO, Kegelmeyer WP (2002) {SMOTE}: {S}ynthetic
	minority over-sampling technique. Journal of Artificial Intelligence Research
	16:321--357, \doi{https://doi.org/10.1613/jair.953}
	
	\bibitem[{Chy and Rahaman(2019)}]{ref6}
	Chy TS, Rahaman MA (2019) A comparative analysis by knn, svm \& elm
	classification to detect sickle cell anemia. In: Proceedings of 2019
	International Conference on Robotics, Electrical and Signal Processing
	Techniques (ICREST 2019), Dhaka, Bangladesh, pp 455--459,
	\doi{10.1109/ICREST.2019.8644410}
	
	\bibitem[{Datta and Das(2015)}]{ref42}
	Datta S, Das S (2015) Near-bayesian support vector machines for imbalanced data
	classification with equal or unequal misclassification costs. Neural Networks
	70:39--52, \doi{10.1016/j.neunet.2015.06.005}
	
	\bibitem[{Doi and Iwasaki(2018)}]{ref47}
	Doi K, Iwasaki A (2018) The effect of focal loss in semantic segmentation of
	high resolution aerial image. In: Proceedings of the 2018 IEEE International
	Geoscience and Remote Sensing Symposium (IGARSS 2018), Valencia, Spain, pp
	6919--6922, \doi{10.1109/IGARSS.2018.8519409}
	
	\bibitem[{Dong  \emph{et~al.}(2015)Dong, Loy, He, and Tang}]{ref52}
	Dong C, Loy CC, He K, Tang X (2015) Image super-resolution using deep
	convolutional networks. IEEE Transactions on Pattern Analysis and Machine
	Intelligence 38(2):295--307, \doi{10.1109/TPAMI.2015.2439281}
	
	\bibitem[{Drummond  \emph{et~al.}(2003)Drummond, Holte  \emph{et~al.}}]{ref51}
	Drummond C, Holte RC,  \emph{et~al.} (2003) C4.5, class imbalance, and cost
	sensitivity: why under-sampling beats over-sampling. In: Proceedings of
	Workshop on Learning from Imbalanced Datasets II, Washington, DC, USA,
	vol~11, pp 1--8
	
	\bibitem[{Durant  \emph{et~al.}(2017)Durant, Olson, Schulz, and Torres}]{ref15}
	Durant TJ, Olson EM, Schulz WL, Torres R (2017) Very deep convolutional neural
	networks for morphologic classification of erythrocytes. Clinical Chemistry
	63(12):1847--1855, \doi{10.1373/clinchem.2017.276345}
	
	\bibitem[{Ford(2013)}]{ref1}
	Ford J (2013) Red blood cell morphology. International Journal of Laboratory
	Hematology 35(3):351--357, \doi{10.1111/ijlh.12082}
	
	\bibitem[{Garc{\i}  \emph{et~al.}(2012)Garc{\i}, Triguero, Carmona, Herrera
		\emph{et~al.}}]{ref35}
	Garc{\i} S, Triguero I, Carmona CJ, Herrera F,  \emph{et~al.} (2012)
	Evolutionary-based selection of generalized instances for imbalanced
	classification. Knowledge-Based Systems 25(1):3--12,
	\doi{10.1016/j.knosys.2011.01.012}
	
	\bibitem[{Garc{\'\i}a and Herrera(2009)}]{ref34}
	Garc{\'\i}a S, Herrera F (2009) Evolutionary undersampling for classification
	with imbalanced datasets: {P}roposals and taxonomy. Evolutionary Computation
	17(3):275--306, \doi{10.1162/evco.2009.17.3.275}
	
	\bibitem[{Garcia  \emph{et~al.}(2007)Garcia, Sanchez, Mollineda, Alejo, and
		Sotoca}]{ref50}
	Garcia V, Sanchez JS, Mollineda RA, Alejo R, Sotoca JM (2007) The class
	imbalance problem in pattern classification and learning. In: Proceedings of
	II Congreso Espanol de Informatica, pp 283--291
	
	\bibitem[{Glorot and Bengio(2010)}]{ref55}
	Glorot X, Bengio Y (2010) Understanding the difficulty of training deep
	feedforward neural networks. In: Proceedings of the 13th International
	Conference on Artificial Intelligence and Statistics (AISTAT 2010), Sardinia,
	Italy, pp 249--256
	
	\bibitem[{Habibzadeh  \emph{et~al.}(2011)Habibzadeh, Krzyzak, and
		Fevens}]{ref4}
	Habibzadeh M, Krzyzak A, Fevens T (2011) Application of pattern recognition
	techniques for the analysis of thin blood smear images. Journal of Medical
	Informatics \& Technologies 18
	
	\bibitem[{Haixiang  \emph{et~al.}(2017)Haixiang, Yijing, Shang, Mingyun,
		Yuanyue, and Bing}]{ref9}
	Haixiang G, Yijing L, Shang J, Mingyun G, Yuanyue H, Bing G (2017) Learning
	from class-imbalanced data: {R}eview of methods and applications. Expert
	Systems with Applications 73:220--239, \doi{10.1016/j.eswa.2016.12.035}
	
	\bibitem[{Han  \emph{et~al.}(2005)Han, Wang, and Mao}]{ref36}
	Han H, Wang WY, Mao BH (2005) {Borderline-SMOTE}: a new over-sampling method in
	imbalanced data sets learning. In: Proceedings of International Conference on
	Intelligent Computing (ICIC 2005), Hefei, China, Springer, pp 878--887,
	\doi{10.1007/11538059_91}
	
	\bibitem[{He  \emph{et~al.}(2008)He, Bai, Garcia, and Li}]{ref38}
	He H, Bai Y, Garcia EA, Li S (2008) {ADASYN}: Adaptive synthetic sampling
	approach for imbalanced learning. In: Proceedings of the IEEE International
	Joint Conference on Neural Networks (IJCNN 2008), Hong Kong, China, pp
	1322--1328, \doi{10.1109/IJCNN.2008.4633969}
	
	\bibitem[{He and Sun(2015)}]{ref56}
	He K, Sun J (2015) Convolutional neural networks at constrained time cost. In:
	Proceedings of the IEEE Conference on Computer Vision and Pattern Recognition
	(CVPR 2015), Boston, MA, USA, pp 5353--5360, \doi{10.1109/CVPR.2015.7299173}
	
	\bibitem[{He  \emph{et~al.}(2016)He, Zhang, Ren, and Sun}]{ref18}
	He K, Zhang X, Ren S, Sun J (2016) Deep residual learning for image
	recognition. In: Proceedings of the IEEE Conference on Computer Vision and
	Pattern Recognition (CVPR 2016), Las Vegas, NV, USA, pp 770--778,
	\doi{10.1109/CVPR.2016.90}
	
	\bibitem[{Hospedales  \emph{et~al.}(2011)Hospedales, Gong, and Xiang}]{ref22}
	Hospedales TM, Gong S, Xiang T (2011) Finding rare classes: {A}ctive learning
	with generative and discriminative models. IEEE Transactions on Knowledge and
	Data Engineering 25(2):374--386, \doi{10.1109/TKDE.2011.231}
	
	\bibitem[{Huang  \emph{et~al.}(2016)Huang, Li, Change~Loy, and Tang}]{ref8}
	Huang C, Li Y, Change~Loy C, Tang X (2016) Learning deep representation for
	imbalanced classification. In: Proceedings of the IEEE conference on computer
	vision and pattern recognition, Las Vegas, NV, USA, pp 5375--5384,
	\doi{10.1109/CVPR.2016.580}
	
	\bibitem[{Huang  \emph{et~al.}(2017)Huang, Liu, Van Der~Maaten, and
		Weinberger}]{ref16}
	Huang G, Liu Z, Van Der~Maaten L, Weinberger KQ (2017) Densely connected
	convolutional networks. In: Proceedings of the IEEE Conference on Computer
	Vision and Pattern Recognition (CVPR 2017), Honolulu, HI, USA, pp 4700--4708,
	\doi{10.1109/CVPR.2017.243}
	
	\bibitem[{Japkowicz and Stephen(2002)}]{ref24}
	Japkowicz N, Stephen S (2002) The class imbalance problem: {A} systematic
	study. Intelligent Data Analysis 6(5):429--449, \doi{10.3233/IDA-2002-6504}
	
	\bibitem[{Kingma and Ba(2014)}]{ref53}
	Kingma DP, Ba J (2014) Adam: A method for stochastic optimization. ArXiv
	Preprint ArXiv:14126980
	
	\bibitem[{Krawczyk(2016)}]{ref26}
	Krawczyk B (2016) Learning from imbalanced data: {O}pen challenges and future
	directions. Progress in Artificial Intelligence 5(4):221--232,
	\doi{10.1007/s13748-016-0094-0}
	
	\bibitem[{LeCun  \emph{et~al.}(2015)LeCun, Bengio, and Hinton}]{ref7}
	LeCun Y, Bengio Y, Hinton G (2015) Deep learning. Nature 521(7553):436,
	\doi{10.1038/nature14539}
	
	\bibitem[{Lee and Chen(2014)}]{ref5}
	Lee H, Chen YPP (2014) Cell morphology based classification for red cells in
	blood smear images. Pattern Recognition Letters 49:155--161,
	\doi{10.1016/j.patrec.2014.06.010}
	
	\bibitem[{Lim  \emph{et~al.}(2016)Lim, Goh, and Tan}]{ref37}
	Lim P, Goh CK, Tan KC (2016) Evolutionary cluster-based synthetic oversampling
	ensemble (eco-ensemble) for imbalance learning. IEEE Transactions on
	Cybernetics 47(9):2850--2861, \doi{10.1109/TCYB.2016.2579658}
	
	\bibitem[{Lin  \emph{et~al.}(2020)Lin, Goyal, Girshick, He, and
		Doll{\'a}r}]{ref10}
	Lin TY, Goyal P, Girshick R, He K, Doll{\'a}r P (2020) Focal loss for dense
	object detection. IEEE Transactions on Pattern Analysis and Machine
	Intelligence 42(2):318--327, \doi{10.1109/TPAMI.2018.2858826}
	
	\bibitem[{Ma  \emph{et~al.}(2018)Ma, Sun, Zhou, Cheng, Chen, and Zhao}]{ref13}
	Ma Y, Sun J, Zhou Q, Cheng K, Chen X, Zhao Y (2018) {CHS-NET}: {A} cascaded
	neural network with semi-focal loss for mitosis detection. In: Proceedings of
	10th Asian Conference on Machine Learning (ACML 2018), Beijing, China., pp
	161--175
	
	\bibitem[{Mazurowski  \emph{et~al.}(2008)Mazurowski, Habas, Zurada, Lo, Baker,
		and Tourassi}]{ref43}
	Mazurowski MA, Habas PA, Zurada JM, Lo JY, Baker JA, Tourassi GD (2008)
	Training neural network classifiers for medical decision making: The effects
	of imbalanced datasets on classification performance. Neural Networks
	21(2-3):427--436, \doi{10.1016/j.neunet.2007.12.031}
	
	\bibitem[{Pasupa and Sunhem(2016)}]{PasupaSunhem:2016}
	Pasupa K, Sunhem W (2016) A comparison between shallow and deep architecture
	classifiers on small dataset. In: Proceeding of the 8th International
	Conference on Information Technology and Electrical Engineering (ICITEE
	2016), Yogyakarta, Indonesia, pp 390--395, \doi{10.1109/ICITEED.2016.7863293}
	
	\bibitem[{Qin  \emph{et~al.}(2018)Qin, Gao, Peng, Wu, Shen, and
		Grudtsin}]{ref45}
	Qin F, Gao N, Peng Y, Wu Z, Shen S, Grudtsin A (2018) Fine-grained leukocyte
	classification with deep residual learning for microscopic images. Computer
	Methods and Programs in Biomedicine 162:243--252,
	\doi{10.1016/j.cmpb.2018.05.024}
	
	\bibitem[{Rahman and Davis(2013)}]{ref25}
	Rahman MM, Davis D (2013) Addressing the class imbalance problem in medical
	datasets. International Journal of Machine Learning and Computing 3(2):224,
	\doi{10.7763/IJMLC.2013.V3.307}
	
	\bibitem[{Razzak and Naz(2017)}]{ref12}
	Razzak MI, Naz S (2017) Microscopic blood smear segmentation and classification
	using deep contour aware {CNN} and extreme machine learning. In: Proceedings
	of 2017 IEEE Conference on Computer Vision and Pattern Recognition Workshops
	(CVPRW 2017), Honolulu, HI, USA, pp 801--807, \doi{10.1109/CVPRW.2017.111}
	
	\bibitem[{Razzak  \emph{et~al.}(2018)Razzak, Naz, and Zaib}]{ref46}
	Razzak MI, Naz S, Zaib A (2018) Deep learning for medical image processing:
	Overview, challenges and the future. In: Classification in BioApps, vol~26,
	pp 323--350, \doi{10.1007/978-3-319-65981-7_12}
	
	\bibitem[{Rehman  \emph{et~al.}(2018)Rehman, Abbas, Saba, Rahman, Mehmood, and
		Kolivand}]{ref21}
	Rehman A, Abbas N, Saba T, Rahman SIu, Mehmood Z, Kolivand H (2018)
	Classification of acute lymphoblastic leukemia using deep learning.
	Microscopy Research and Technique 81(11):1310--1317, \doi{10.1002/jemt.23139}
	
	\bibitem[{Ross  \emph{et~al.}(2006)Ross, Pritchard, Rubin, and Duse}]{ref14}
	Ross NE, Pritchard CJ, Rubin DM, Duse AG (2006) Automated image processing
	method for the diagnosis and classification of malaria on thin blood smears.
	Medical and Biological Engineering and Computing 44(5):427--436,
	\doi{10.1007/s11517-006-0044-2}
	
	\bibitem[{Russakovsky  \emph{et~al.}(2015)Russakovsky, Deng, Su, Krause,
		Satheesh, Ma, Huang, Karpathy, Khosla, Bernstein  \emph{et~al.}}]{ref19}
	Russakovsky O, Deng J, Su H, Krause J, Satheesh S, Ma S, Huang Z, Karpathy A,
	Khosla A, Bernstein M,  \emph{et~al.} (2015) Imagenet large scale visual
	recognition challenge. International Journal of Computer Vision
	115(3):211--252, \doi{10.1007/s11263-015-0816-y}
	
	\bibitem[{Srivastava  \emph{et~al.}(2015)Srivastava, Greff, and
		Schmidhuber}]{ref54}
	Srivastava RK, Greff K, Schmidhuber J (2015) Highway networks. ArXiv Preprint
	ArXiv:150500387
	
	\bibitem[{Sudre  \emph{et~al.}(2017)Sudre, Li, Vercauteren, Ourselin, and
		Cardoso}]{ref29}
	Sudre CH, Li W, Vercauteren T, Ourselin S, Cardoso MJ (2017) Generalised dice
	overlap as a deep learning loss function for highly unbalanced segmentations.
	In: Proceedings of the 3rd International Workshop, Deep Learning in Medical
	Image Analysis (DLMIA 2017) and 7th International Workshop in Multimodal
	Learning for Clinical Decision Support (CDS 2017), QC, Canada, pp 240--248,
	\doi{10.1007/978-3-319-67558-9_28}
	
	\bibitem[{Suresh  \emph{et~al.}(2008)Suresh, Sundararajan, and
		Saratchandran}]{ref32}
	Suresh S, Sundararajan N, Saratchandran P (2008) Risk-sensitive loss functions
	for sparse multi-category classification problems. Information Sciences
	178(12):2621--2638, \doi{10.1016/j.ins.2008.02.009}
	
	\bibitem[{Taherisadr  \emph{et~al.}(2013)Taherisadr, Nasirzonouzi, Baradaran,
		and Mehdizade}]{ref3}
	Taherisadr M, Nasirzonouzi M, Baradaran B, Mehdizade A (2013) New approch to
	red blood cell classification using morphological image processing. Shiraz
	E-Medical Journal 14(1):44--53
	
	\bibitem[{Tian  \emph{et~al.}(2018)Tian, Wu, Wang, and Cao}]{ref48}
	Tian X, Wu D, Wang R, Cao X (2018) Focal text: an accurate text detection with
	focal loss. In: Proceedings of the 25th IEEE International Conference on
	Image Processing (ICIP 2018), Athens, Greece, pp 2984--2988,
	\doi{10.1109/ICIP.2018.8451241}
	
	\bibitem[{Tiwari  \emph{et~al.}(2018)Tiwari, Qian, Li, Wang, Gupta, Khanna,
		Rodrigues, and de~Albuquerque}]{ref44}
	Tiwari P, Qian J, Li Q, Wang B, Gupta D, Khanna A, Rodrigues JJ, de~Albuquerque
	VHC (2018) Detection of subtype blood cells using deep learning. Cognitive
	Systems Research 52:1036--1044, \doi{10.1016/j.cogsys.2018.08.022}
	
	\bibitem[{Tomari  \emph{et~al.}(2014)Tomari, Zakaria, Jamil, Nor, and
		Fuad}]{ref2}
	Tomari R, Zakaria WNW, Jamil MMA, Nor FM, Fuad NFN (2014) Computer aided system
	for red blood cell classification in blood smear image. Procedia Computer
	Science 42:206--213, \doi{10.1016/j.procs.2014.11.053}
	
	\bibitem[{Wang  \emph{et~al.}(2016)Wang, Liu, Wu, Cao, Meng, and
		Kennedy}]{ref30}
	Wang S, Liu W, Wu J, Cao L, Meng Q, Kennedy PJ (2016) Training deep neural
	networks on imbalanced data sets. In: Proceedings of 2016 International Joint
	Conference on Neural Networks (IJCNN 2016), Vancouver, BC, Canada, pp
	4368--4374, \doi{10.1109/IJCNN.2016.7727770}
	
	\bibitem[{Weiss(2004)}]{ref23}
	Weiss GM (2004) Mining with rarity: {A} unifying framework. ACM SIGKDD
	Explorations Newsletter 6(1):7--19, \doi{10.1145/1007730.1007734}
	
	\bibitem[{Xu  \emph{et~al.}(2017)Xu, Papageorgiou, Abidi, Dao, Zhao, and
		Karniadakis}]{ref11}
	Xu M, Papageorgiou DP, Abidi SZ, Dao M, Zhao H, Karniadakis GE (2017) A deep
	convolutional neural network for classification of red blood cells in sickle
	cell anemia. PLoS Computational Biology 13(10):e1005746,
	\doi{10.1371/journal.pcbi.1005746}
	
	\bibitem[{Yue(2017)}]{ref28}
	Yue S (2017) Imbalanced malware images classification: {A} {CNN} based
	approach. ArXiv Preprint ArXiv:170808042
	
	\bibitem[{Zhang  \emph{et~al.}(2018)Zhang, Tan, Li, and Hong}]{ref27}
	Zhang C, Tan KC, Li H, Hong GS (2018) A cost-sensitive deep belief network for
	imbalanced classification. IEEE Transactions on Neural Networks and Learning
	Systems 30(1):109--122, \doi{10.1109/TNNLS.2018.2832648}
	
	\bibitem[{Zhang  \emph{et~al.}(2017)Zhang, Fang, Wen, Li, and Qiao}]{ref31}
	Zhang X, Fang Z, Wen Y, Li Z, Qiao Y (2017) Range loss for deep face
	recognition with long-tailed training data. In: Proceedings of the IEEE
	International Conference on Computer Vision (ICCV 2017), Venice, Italy, pp
	5409--5418, \doi{10.1109/ICCV.2017.578}
	
	\bibitem[{Zhao  \emph{et~al.}(2017)Zhao, Zhang, Zhou, Chu, and Cao}]{ref20}
	Zhao J, Zhang M, Zhou Z, Chu J, Cao F (2017) Automatic detection and
	classification of leukocytes using convolutional neural networks. Medical \&
	Biological Engineering \& Computing 55(8):1287--1301,
	\doi{10.1007/s11517-016-1590-x}
	
	\bibitem[{Zhou and Liu(2006)}]{ref39}
	Zhou ZH, Liu XY (2006) Training cost-sensitive neural networks with methods
	addressing the class imbalance problem. IEEE Transactions on Knowledge \&
	Data Engineering 18(1):63--77, \doi{10.1109/TKDE.2006.17}
	
	\bibitem[{Zong  \emph{et~al.}(2013)Zong, Huang, and Chen}]{ref40}
	Zong W, Huang GB, Chen Y (2013) Weighted extreme learning machine for imbalance
	learning. Neurocomputing 101:229--242, \doi{10.1016/j.neucom.2012.08.010}
	
\end{thebibliography}


\end{document}